\pdfoutput=1
\documentclass[aps,prd,preprint,superscriptaddress,showpacs]{revtex4}
\usepackage{graphicx}
\usepackage{amsmath}
\usepackage{bm}% bold math
\usepackage{amssymb,enumerate}
\usepackage{subfigure}

\def\ee{\end{eqnarray}}

\def\p{\partial}

\def\sech{\mathop{\mathrm{sech}}\nolimits}

\def\=:{=\hspace{-.7em}\raisebox{1.1ex}{.}\hspace{.1em}\raisebox{-0.2ex}{.} }
\def\ee{\end{eqnarray}}

\def\p{\partial}

\def\sech{\mathop{\mathrm{sech}}\nolimits}

\def\=:{=\hspace{-.7em}\raisebox{1.1ex}{.}\hspace{.1em}\raisebox{-0.2ex}{.} }

\newcommand {\beq}{\begin{eqnarray}}
\newcommand {\eeq}{\end{eqnarray}}
\newcommand {\non}{\nonumber\\}

\newcommand {\1}[1]{\frac{1}{#1}}

\newcommand {\del}{\partial}

\newcommand {\tr}{{\rm tr}\,}

\begin{document}

% Use the \preprint command to place your local institutional report
% number in the upper righthand corner of the title page in preprint mode.
% Multiple \preprint commands are allowed.
% Use the 'preprintnumbers' class option to override journal defaults
% to display numbers if necessary
\preprint{\tt NORDITA-2014-90}

%Title of paper
\title{Incarnations of Skyrmions
}

% repeat the \author .. \affiliation  etc. as needed
% \email, \thanks, \homepage, \altaffiliation all apply to the current
% author. Explanatory text should go in the []'s, actual e-mail
% address or url should go in the {}'s for \email and \homepage.
% Please use the appropriate macro foreach each type of information

% \affiliation command applies to all authors since the last
% \affiliation command. The \affiliation command should follow the
% other information
% \affiliation can be followed by \email, \homepage, \thanks as well.
\author{Sven Bjarke Gudnason}
\affiliation{{\it Nordita, KTH Royal Institute of Technology and
    Stockholm University, Roslagstullsbacken 23, SE-106 91 Stockholm,
    Sweden }}
\author{Muneto  Nitta}
\affiliation{{\it Department of Physics, and Research and
    Education Center for Natural Sciences, Keio University, Hiyoshi
    4-1-1, Yokohama, Kanagawa 223-8521, Japan}}
%Collaboration name if desired (requires use of superscriptaddress
%option in \documentclass). \noaffiliation is required (may also be
%used with the \author command).
%\collaboration can be followed by \email, \homepage, \thanks as well.
%\collaboration{}
%\noaffiliation

%Collaboration name if desired (requires use of superscriptaddress
%option in \documentclass). \noaffiliation is required (may also be
%used with the \author command).
%\collaboration can be followed by \email, \homepage, \thanks as well.
%\collaboration{}
%\noaffiliation

\date{\today}
\begin{abstract}
Skyrmions can be transformed into lumps or baby-Skyrmions 
by being trapped inside a domain wall. 
Here we find that they can also be transformed into sine-Gordon kinks 
when confined by vortices, resulting in confined Skyrmions.
We show this both by an effective field theory approach and by direct
numerical calculations. 
The existence of these trapped and confined Skyrmions does not rely on
higher-derivative terms when the host solitons are flat or straight. 
We also construct a Skyrmion as a twisted vortex ring in a model with
a sixth-order derivative term. 

\end{abstract}
% insert suggested PACS numbers in braces on next line
% insert suggested keywords - APS authors don't need to do this
%\keywords{}

%\maketitle must follow title, authors, abstract, \pacs, and \keywords
\maketitle

\section{Introduction}

Topological solitons and instantons \cite{Manton:2004tk} play a
significant role in diverse areas of physics such as quantum field
theories, string theory, cosmology \cite{Vilenkin:2000} and condensed
matter systems \cite{Volovik:2003}. 
For instance, Yang-Mills instantons play an especially important
role in non-perturbative dynamics of quantum gauge-field theories. 
Relations between solitons and instantons with different
dimensionalities are important for a unified understanding 
of these objects, which may explain or even reveal unexpected
relations between field theories defined in different dimensions. 
Yang-Mills instantons in pure Yang-Mills theory in four-dimensional
Euclidean space are particle-like topological solitons in $d=4+1$
dimensional spacetime. 
When coupled to Higgs fields in the Higgs phase, they are unstable and 
shrink in the bulk. 
They can, however, stably live inside host solitons, in which they
transform themselves to other kinds of solitons:
They transform themselves into Skyrmions \cite{Eto:2005cc} when
trapped inside a non-Abelian domain wall \cite{Shifman:2003uh}, or
lumps \cite{Eto:2004rz} when inside a non-Abelian vortex
\cite{Hanany:2003hp,Eto:2005yh} and sine-Gordon (SG) kinks
\cite{Nitta:2013cn} when inside a monopole string, as summarized in
(d), (e), and (f) in Tab.~\ref{table:instantons}. 
One such composites, namely a lump inside a non-Abelian vortex, 
elegantly explains the coincidence of mass spectra between 
field theories in different dimensions
\cite{Dorey:1998yh,Shifman:2004dr}. 
If one compactifies the world volume of host solitons, instantons also
can exist, not only as such trapped solitons; they become twisted
closed domain walls, vortex sheets or monopole strings, 
when the moduli $S^3$, $S^2$, or $S^1$ of these host solitons are 
wound around their compact world volumes, also of the shape
$S^3$, $S^2$, or $S^1$, respectively \cite{Nitta:2013vaa} 
[see (d), (e), and (f) in Tab.~\ref{table:instantons}]. 

%%%%%%%%%%%%%%%%%%%%%%%%%
\begin{table}[!htb]
\begin{tabular}{|c|c||c|c|c|c|c|c|c|c|} \hline
 soliton & $\pi_n$ in & & host solitons &  $\pi_n$ of & codim &  moduli &  w.v. &  w.v. & $\pi_n$ on \\ 
 /dim & bulk& &  & host  &  & &shape  & soliton  & w.v.\\ \hline\hline
 lump & $\pi_2(S^2)$ &  (a) & $ {\mathbb C}P^1$ domain wall& $\pi_0$ & $1$ & $S^1$ & ${\mathbb R}^1$ or $S^1$ & SG kink & $\pi_1(S^1)$ \\ 
 2+1 dim & & & & & & & & & \\
\hline
%%%
 Skyrmion & $\pi_3(S^3)$ &(b) & NA domain wall & $\pi_0$ & $1$  & $S^2$ &  ${\mathbb R}^2$ or $S^2$ & lump &$\pi_2(S^2)$\\
3+1 dim & & (c) & vortex string & $\pi_1$ & $2$  & $S^1$ &   ${\mathbb R}^1$ or $S^1$ & SG kink & $\pi_1(S^1)$\\ \hline
%%%
Instanton & $\pi_3(G)$ &(d) &NA domain wall& $\pi_0$ & $1$ & $S^3$ & ${\mathbb R}^3$ or 
$S^3$ & Skyrmion & $\pi_3(S^3)$ \\
4+1 dim & & (e) &NA vortex sheet & $\pi_1$  &  $2$  & $S^2$ &  ${\mathbb R}^2$ or $S^2$ & lump &$\pi_2(S^2)$\\
 & & (f) &monopole string & $\pi_2$ & $3$  & $S^1$ &   ${\mathbb R}^1$ or $S^1$ & SG kink & $\pi_1(S^1)$\\ \hline
\end{tabular}
\caption{Host solitons of 
trapped instantons (Skyrmions). 
{\small (a), (d), (e) and (f)  were already pointed out in
  Ref.~\cite{Nitta:2013vaa}. 
  The shape of the world-volume can be noncompact, ${\mathbb R}^n$,
  or compact, $S^n$, corresponding to trapped and untrapped instantons
  (Skyrmions), respectively. 
  w.v.~stands for ``world volume,'' NA denotes ``non-Abelian'' and $G$
  denotes the gauge group. } 
\label{table:instantons}}
\end{table}
%%%%%%%%%%%%%%%%%%%%%%

A similar relation is known in $d=2+1$ dimensions, in which lumps
\cite{Polyakov:1975yp} or baby-Skyrmions
\cite{Piette:1994ug,Weidig:1998ii} become sine-Gordon kinks
\cite{Nitta:2012xq,Kobayashi:2013ju,Jennings:2013aea} when trapped
inside a ${\mathbb C}P^1$ domain wall
\cite{Abraham:1992vb,Arai:2002xa}, corresponding to (a) in
Tab.~\ref{table:instantons}. 
This relation was already known earlier in condensed-matter systems
such as Josephson junctions of two superconductors
\cite{Ustinov:1998}, ferromagnets \cite{Chen:1977}, and $^3$He 
superfluids \cite{Volovik:2003}. 
When one compactifies the world volume of the domain wall to $S^1$, it
becomes an isolated lump or baby-Skyrmion as a closed domain line with
a twisted $U(1)$ modulus \cite{Kobayashi:2013ju}. 
Similar twisted closed wall lines as vortices also exist in
condensed-matter systems such as p-wave superconductors
\cite{Garaud:2012}.

\begin{figure}[!tp]
\begin{center}
\mbox{
\subfigure[]{\includegraphics[width=0.25\linewidth]{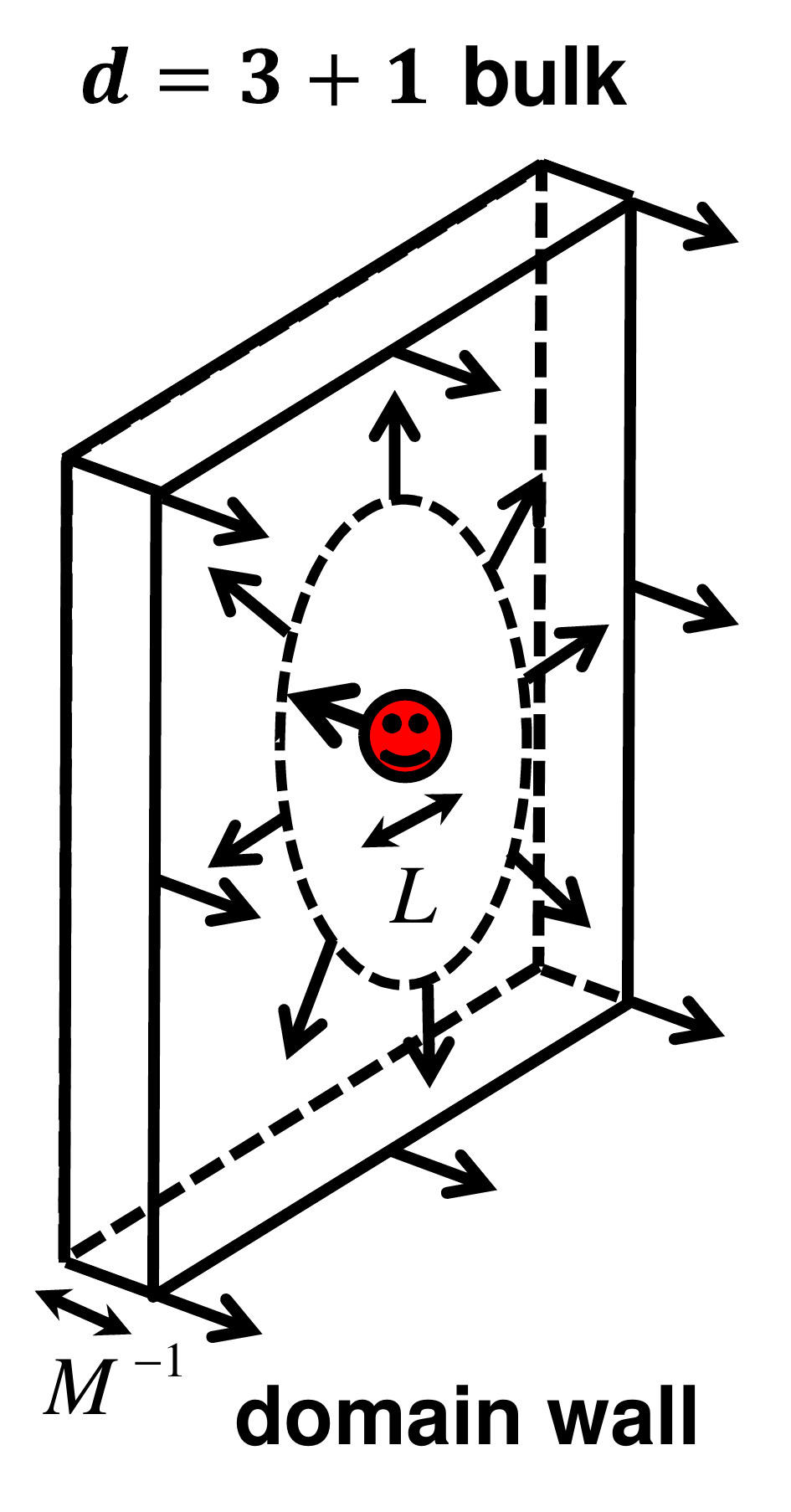}}\qquad\qquad
\subfigure[]{\includegraphics[width=0.2\linewidth]{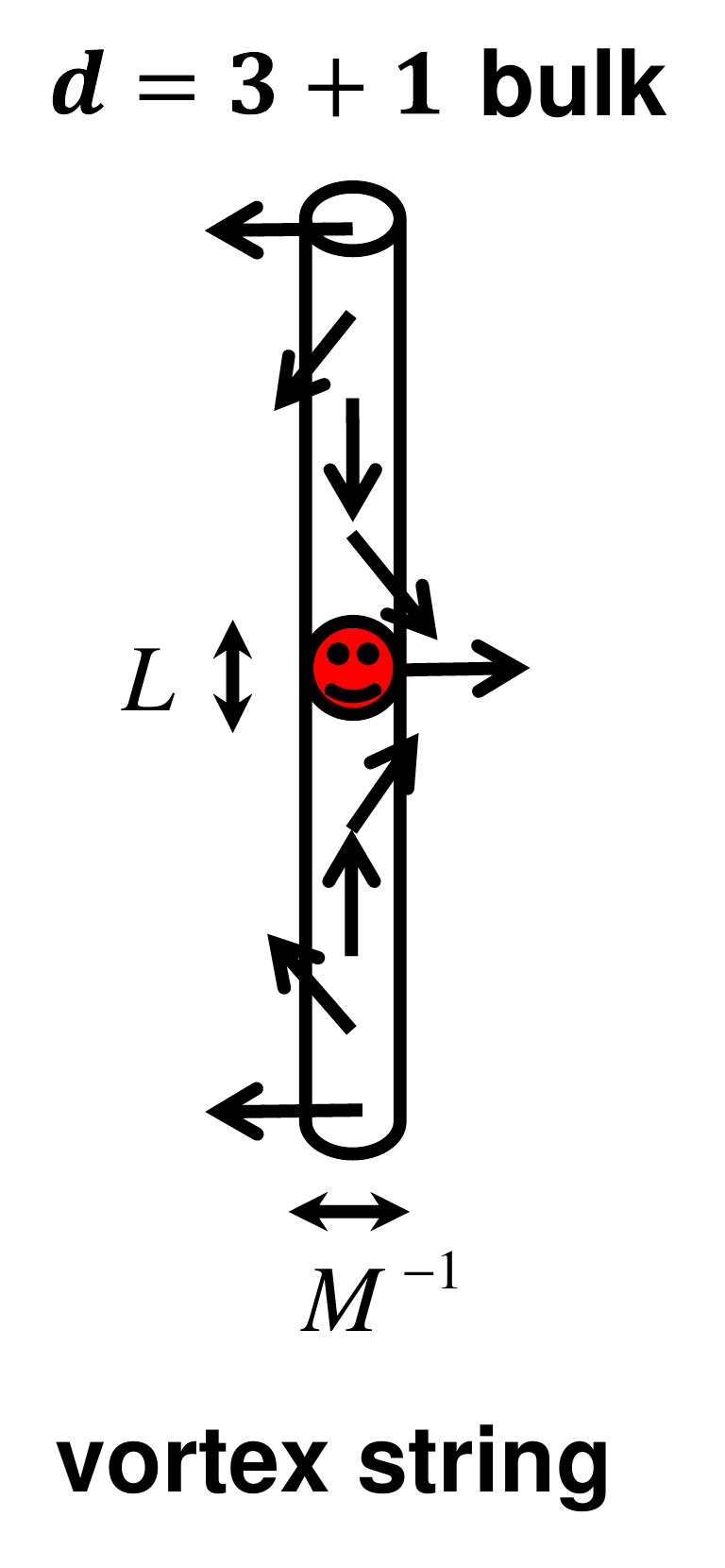}}}
\mbox{
\subfigure[]{\includegraphics[width=0.3\linewidth]{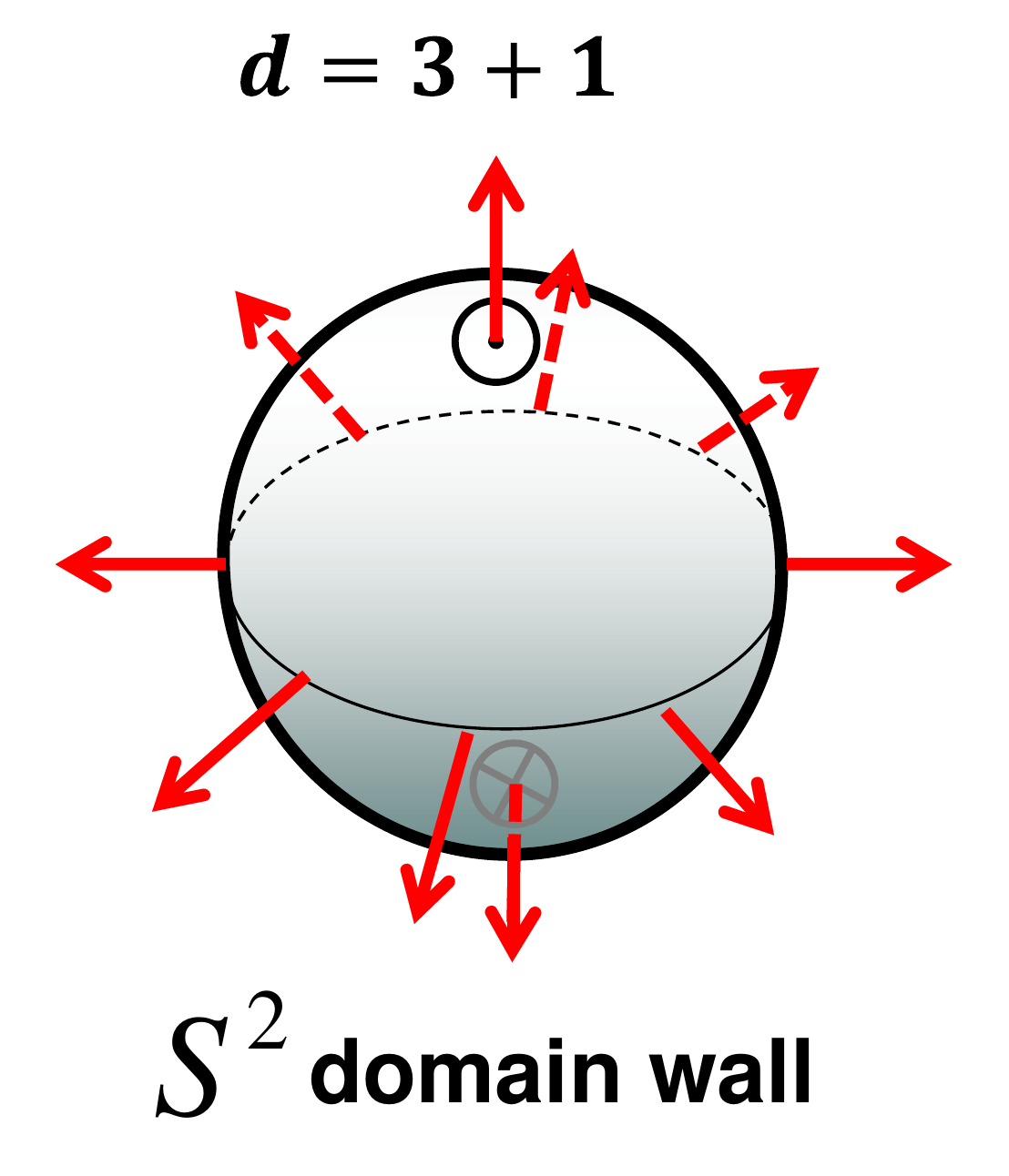}}\quad
\subfigure[]{\includegraphics[width=0.3\linewidth]{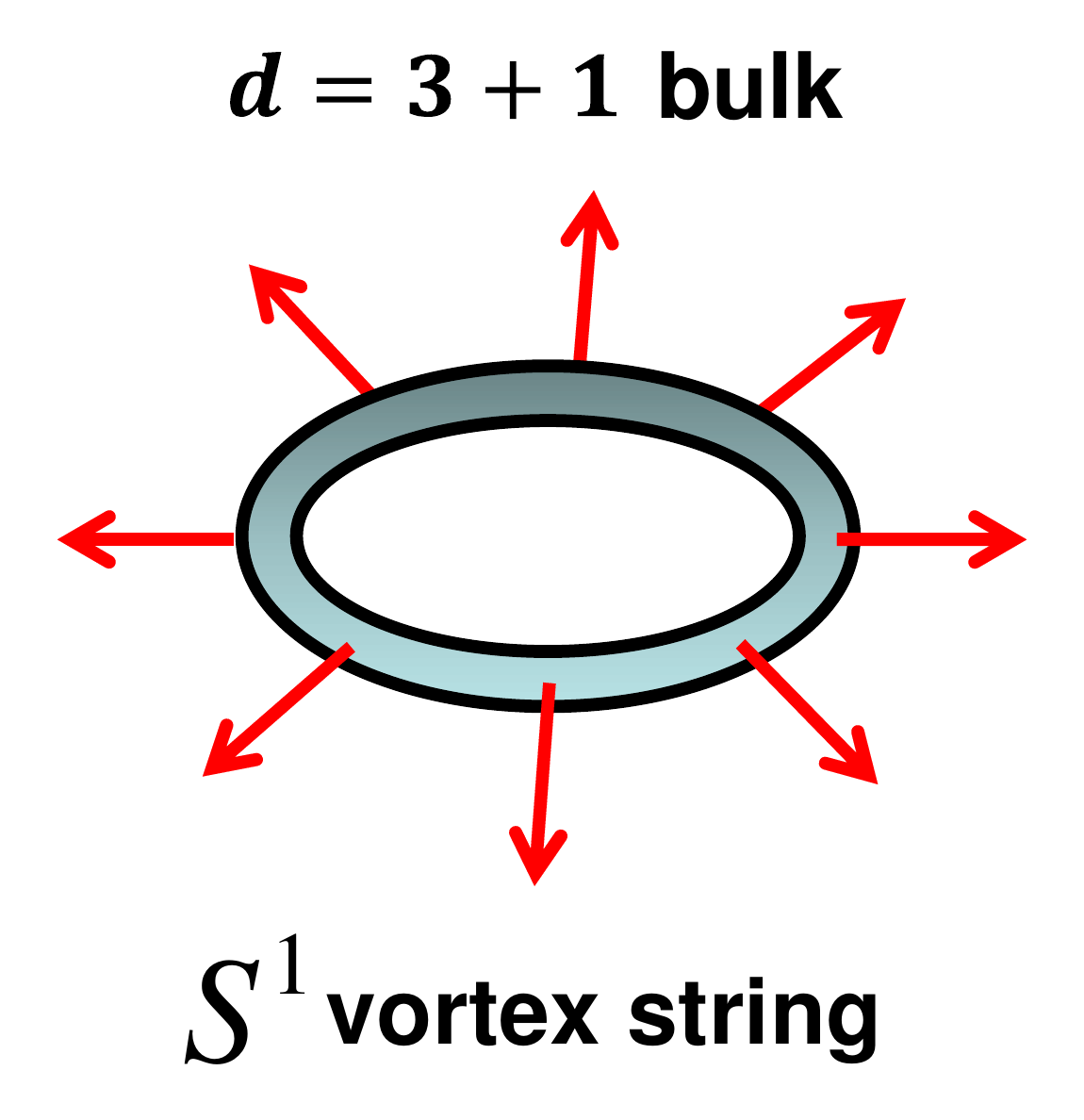}}
}
\caption{Incarnation of Skyrmions.  
  (a) A lump or a baby-Skyrmion on a flat domain wall,
  (b) A sine-Gordon kink on a straight vortex, 
  (c) A spherical domain wall with twisted $S^2$ moduli, 
  (d) A vortex ring with twisted $S^1$ modulus.
\label{fig:incarnation}}
\end{center}
\end{figure}

In this paper, we further pursue these relations between solitons with
different dimensionalities, by focusing on Skyrmions
\cite{Skyrme:1962vh} in $d=3+1$ dimensions, which are characterized by
the topological charge $\pi_3(S^3) \simeq {\mathbb Z}$, i.e.~the
baryon number.
It was already found that Skyrmions become baby-Skyrmions or lumps
when trapped inside a non-Abelian domain wall
\cite{Nitta:2012wi,Nitta:2012rq,Gudnason:2014nba}, and a spherical
domain wall with twisted $S^2$ moduli is a Skyrmion
\cite{Gudnason:2013qba}, both corresponding to (b) in
Tab.~\ref{table:instantons}  
(this relation was generalized to $N$-dimensional Skyrmions becoming
$N-1$ dimensional Skyrmions inside non-Abelian domain walls in
\cite{Nitta:2012rq}). 
The last piece which was missing in Tab.~\ref{table:instantons}
corresponds to (c), which we will work out explicitly in this paper. 
We consider a potential term motivated by two-component Bose-Einstein
condensates (BECs) of ultracold atoms with repulsive interactions
\cite{Kasamatsu:2005} (see Appendix \ref{app:BEC}), 
which are known to admit a variety of topological 
solitons such as domain walls, vortices, 
Skyrmions (as vortons)
\cite{3D-skyrmions,Nitta:2012hy,Metlitski:2003gj} 
and D-brane solitons (vortices ending on a domain wall)
\cite{Kasamatsu:2010aq,Nitta:2012hy}.  
We denote this model the BEC Skyrme model.
The model admits a global vortex solution with a $U(1)$ modulus.
We then deform the potential by a perturbation which introduces a
potential for the $U(1)$ modulus, and construct a stable Skyrmion as a
sine-Gordon kink trapped inside the straight vortex. 
We achieve this by two approaches: the effective field theory on
solitons \cite{Gudnason:2014gla} which is based on the moduli
approximation \cite{Manton:1981mp} and by direct numerical
computations. 
Together with the previous result, we have two kinds of solitons:
a domain wall and a vortex, able to host Skyrmions, 
as illustrated in Fig.~\ref{fig:incarnation} (a) and (b). 

%%%%%%%%%%%%%%%%%%%%%%
%%%%%%%%%%%%%%%%%%%%%%

We then compactify the world-volume of the vortex-line to $S^1$ for
which we consider the BEC potential without any perturbations. 
The configuration becomes a vortex ring with a twisted $U(1)$
modulus, that is, a vorton \cite{Davis:1988jq} (except for the time
dependence which is usually required for vortons).  
In fact, it is known in the context of BECs that a Skyrmion is nothing 
but a vorton \cite{3D-skyrmions,Nitta:2012hy,Metlitski:2003gj}.
Together with the previous results of \cite{Gudnason:2014nba}, we find
two possible incarnations of Skyrmions; one corresponds to a Skyrmion
as a spherical domain wall with the $S^2$ moduli twisted along the
$S^2$ world-volume \cite{Gudnason:2013qba}, while the other
corresponds to a closed vortex string with the $U(1)$ modulus twisted
along the $S^1$ world-volume, as illustrated in
Fig.~\ref{fig:incarnation} (c) and (d). 

For a certain choice of perturbed potential which is described above,
a Skyrmion is attached by the same kind of vortices from both of its
sides. 
With a different choice of perturbed potential term, we also construct
a (half-)Skyrmion attached by different kinds of vortices from both of
its sides. 
In this case, we cannot compactify the world-volume since the vortices
attached from both of its sides are different.  
The latter is similar to a confined monopole in the Higgs phase:
In the Higgs phase, magnetic fluxes of a 't Hooft-Polyakov monopole
\cite{'tHooft:1974qc} are squeezed to form vortices, and the monopole
becomes a kink inside a vortex
\cite{Tong:2003pz,Shifman:2004dr,Nitta:2010nd}.  
This was the first prime example of composite 
Bogomol'nyi-Prasad-Sommerfield (BPS) solitons, 
see Refs.~\cite{Tong:2005un,Eto:2006pg,Shifman:2007ce} for a review. 
We call our configurations confined Skyrmions. 

This paper is organized as follows.
In Sec.~\ref{sec:model}, we present the Skyrme models we consider in
this paper. In particular, we use two different kinds of potential
terms for model 1 and 2, respectively. 
Model 1 was already studied in our previous works, except for the
lump-type of Skyrmion residing in the domain wall, while model 2 is
introduced in this paper and is motivated by two-component BECs. 
In Sec.~\ref{sec:wall-vortex}, we construct domain wall solutions for
model 1 and global vortex solutions for model 2, which will serve as
host solitons for our baby solitons: baby Skyrmions (or lumps) and
sine-Gordon kinks, respectively.
In Sec.~\ref{sec:effective}, we construct Skyrmions being
baby-Skyrmions (or lumps) and sine-Gordon kinks in the effective
theories of the domain wall and the vortex as host solitons in model 1
and model 2, respectively. 
In Sec.~\ref{sec:numerical}, we provide full numerical solutions for 
such composite Skyrmions.  
We also construct a vortex ring with the twisted $U(1)$ modulus as a
Skyrmion in model 2. 
Sec.~\ref{sec:summary} is devoted to a summary and discussion.
In Appendix \ref{app:BEC}, we explain a potential term of
two-component BECs and its relation to our model. 
In Appendix \ref{app:half}, we show full numerical solutions of a
half-Skyrmion trapped inside a vortex in model 2.

\newpage 
%%%%%%%%%%%%%%%%%%%%%%%%%%%%%%%%%%
\section{Skyrme-like models}
\label{sec:model}

We consider the $SU(2)$ principal chiral model or the Skyrme model 
with higher-derivative terms, in $d=3+1$ dimensions. 
With the $SU(2)$ valued field $U(x)\in SU(2)$, the Lagrangian which we 
consider is given by 
\beq
\mathcal{L} = \frac{f_{\pi}^2}{16} \tr (\del_{\mu}U^{\dagger}
\del^{\mu} U) + \mathcal{L}_4 + \mathcal{L}_6 - V(U),
\eeq
with the Skyrme \cite{Skyrme:1962vh} and sixth-order derivative term,
respectively 
\begin{align}  
\mathcal{L}_4 &= \frac{\kappa}{32e^2} \tr 
\left([U^\dagger \del_{\mu} U, U^\dagger \del_{\nu} U]^2\right),
\label{eq:Skyrme-term}\\
\mathcal{L}_6 &=
\frac{c_6}{36 e^4 f_\pi^2}\left(\epsilon^{\mu\nu\rho\sigma} 
\tr\left[U^\dagger\del_\nu U U^\dagger\del_\rho U 
  U^\dagger\del_\sigma U\right]\right)^2.
\label{eq:6th_order-term}
\end{align}
The symmetry of the Lagrangian with $V=0$ is 
$\tilde G = SU(2)_{L} \times SU(2)_{R}$ acting on $U$ as 
$U \to U'= g_L U g_R^\dagger$.
This is spontaneously broken to $\tilde H \simeq SU(2)_{L+R}$
acting as 
$U \to U'= g_{L+R} U g_{L+R}^\dagger$ so that the target space is 
$\tilde G/\tilde H \simeq SU(2)_{\rm L-R}$.
The conventional potential term is 
$V = m_{\pi}^2\tr(2{\bf 1}_2 - U - U^\dagger)$,
which breaks the symmetry $\tilde{G}$ to $SU(2)_{\rm L+R}$
\emph{explicitly}.

In this paper, we consider both cases where the higher-derivative
terms are turned off ($\kappa=c_6=0$) and where either the Skyrme term
or the sixth-order term is turned on. 
A BPS model was discovered some years back \cite{Adam:2010fg},
which consists of only the sixth-order term as well as appropriate
potentials. This type of model is not in our parameter space here as
it corresponds to $f_\pi\to 0$ (and $\kappa=0$) which is not possible
as we will rescale away $f_\pi$, see below. 
It is interesting, however, for phenomenological reasons due to the
possibility of parametrically small (classically) binding energies and
a large extended symmetry (volume-preserving diffeomorphisms). In this
paper, we consider systems which perhaps are closer to
condensed-matter systems and do have a kinetic term and large binding
energies here are thus not an immediate concern.  

If we rewrite the Lagrangian in terms of two complex scalar fields 
$\phi^{\rm T}=\{\phi_1(x),\phi_2(x)\}$, defined by
\beq
U = 
\begin{pmatrix}
\phi_1 & -\phi_2^*\\
\phi_2 & \phi_1^*
\end{pmatrix},
\eeq
subject to the constraint
\beq
\det U = |\phi_1|^2 + |\phi_2|^2 = 1,
\eeq
then the Lagrangian can be rewritten as 
\begin{align}
\mathcal{L} 
&= \1{2} \partial_{\mu} \phi^\dagger \partial^{\mu}\phi       
- \frac{\kappa}{4}
\left[ (\partial_{\mu} \phi^\dagger \partial^{\mu}\phi )^2  
-\1{4}(\partial_{\mu} \phi^\dagger \partial_{\nu}\phi + 
\partial_{\nu} \phi^\dagger \partial_{\mu}\phi )^2
\right] \non
&\phantom{=\ } 
+ \frac{c_6}{144}
\left(\epsilon^{\mu\nu\rho\sigma} \left[
  \phi^\dagger \left(\del_\nu\phi\del_j\phi^\dagger +
  \sigma^2\del_i\phi^*\del_\rho\phi^{\rm T}\sigma^2\right)\del_\sigma\phi +
        {\rm c.c.}\right]\right)^2
- V(\phi,\phi^*) \non
&= \1{2} \partial_{\mu}  {\bf n}  \cdot \partial^{\mu}  {\bf n}       
- \frac{\kappa}{4}
\left[ (\partial_{\mu}  {\bf n}  \cdot \partial^{\mu}  {\bf n} )^2  
 - (\partial_{\mu} {\bf n} \cdot \partial_{\nu} {\bf n}  )^2
\right] \non
&\phantom{=\ } +
\frac{c_6}{36}\left(\epsilon^{\mu\nu\rho\sigma}\epsilon^{ABCD}\del_\nu
n_A \del_\rho n_B \del_\sigma n_C n_D\right)^2 
- V( {\bf n} ), 
\label{eq:masterL_dimless}
\end{align}
where we have rescaled the Lagrangian density such that energy is
measured in units of $f_\pi/(2e)$ and length is measured in units of
$2/(ef_\pi)$ and we have introduced the real four-vector scalar fields 
${\bf n}(x)=\{n_A(x)\}=\{n_1(x),n_2(x),n_3(x),n_4(x)\}$ satisfying 
${\bf n}^2 =\sum_A n_A^2 = 1$, 
defined by 
$\phi_1 = n_1 + i n_2$ and $\phi_2 = n_3 + i n_4$ 
($A,B,C,D=1,2,3,4$).

The target space (the vacuum manifold with $V=0$) 
$\mathcal{M}\simeq SU(2)\simeq S^3$ has a nontrivial homotopy group 
\beq
\pi_3(\mathcal{M}) = {\mathbb Z}, 
\eeq
which admits Skyrmions as usual. 
The baryon number (Skyrme charge), $B \in \pi_3(S^3)$, is defined as 
\begin{align}
B &= -\1{24\pi^2} \int d^3x \;
\epsilon^{ijk} \, \tr \left( U^\dagger \partial_i U 
             U^\dagger \partial_j U U^\dagger \partial_k U\right) \non
&= \1{24\pi^2} \int d^3x \;
 \epsilon^{ijk} \, \tr \left( U^\dagger \partial_i U 
           \partial_j U^\dagger \partial_k U\right) \non
&= \frac{1}{24 \pi^2}  \int d^3x \left[\epsilon^{ijk} 
  \phi^\dagger \left(\del_i\phi\del_j\phi^\dagger +
  \sigma^2\del_i\phi^*\del_j\phi^{\rm T}\sigma^2\right)\del_k\phi +
        {\rm c.c.}\right] \non
&= -\frac{1}{12 \pi^2} \int d^3x\; \epsilon^{ABCD} \epsilon^{ijk} 
  \partial_i n_A \partial_j n_B \partial_k n_C n_D \non
&= -\frac{1}{2\pi^2} \int d^3x\; \epsilon^{ABCD} 
  \partial_1 n_A \partial_2 n_B \partial_3 n_C n_D.
\end{align}

Instead of the conventional potential term, we consider here potential  
terms of the form 
\beq
V = V_1 + V_2,
\eeq
where $V_1$ is the dominant potential and it admits a host soliton
such as a domain wall or a vortex while $V_2$ is a subdominant
potential admitting a soliton inside of the host soliton -- a baby 
soliton. 
We consider two theories: model 1 admitting Skyrmions as
baby-Skyrmions inside a domain wall; and model 2 admitting Skyrmions
as sine-Gordon solitons inside a vortex.

For model 1, we take the potential to be
\beq
 V_1 =  \frac{1}{2}M^2 (1-n_4^2), \qquad
 V_2 = -\frac{1}{2}m_3^2 n_3^{a_3}
\eeq
with $a_3=1$ or $2$. 
When $m_3$ is zero, $V_2$ vanishes and the potential admits two
discrete vacua: $n_4 = \pm 1$.
This allows for a domain wall interpolating between the two vacua 
\cite{Kudryavtsev:1999zm} 
as we show in the next section. 
With a nonzero $m_3>0$, there still remain two vacua and a domain wall
interpolating between them as long as $m_3<M$ 
\cite{Nitta:2012wi,Nitta:2012rq,Gudnason:2014nba}. 

For model 2, we take the potential to be 
\begin{align}
V_1 &=\frac{1}{8}M^2\left[1-\big(\phi^\dag\sigma^3\phi\big)^2\right] 
= \frac{1}{2} M^2 |\phi_1|^2|\phi_2|^2 
= \frac{1}{2} M^2 \left(n_1^2 + n_2^2\right) 
\left(n_3^2 + n_4^2\right), \non
V_2 &= - \frac{1}{2}m_3^2 n_3^{a_3},
\end{align}
with $a_3=1$ or $2$.
The potential $V_1$ is motivated by two-component BECs 
(see Appendix \ref{app:BEC}), and admits
global vortices \cite{Kasamatsu:2005} 
and Skyrmions as vortons
\cite{3D-skyrmions,Nitta:2012hy,Metlitski:2003gj}.

%%%%%%%%%%%%%%%%%%%%%%%%%%%%%%%%%%
\section{Domain walls and vortices as host solitons}
\label{sec:wall-vortex}

In this section, we construct a domain wall and a vortex for models 1
and 2, respectively, with their respective $V_1$ potentials in the
limit $V_2=0$. 
We will take into account the effect of $V_2$ in the next
sections. 

%%%%%%%
\subsection{Model 1: the domain wall}
\label{sec:model1}

Model 1 has two discrete vacua and admits a domain wall solution
interpolating between them. 
First, we consider only $V_1$ with $m_3=0$ ($V_2=0$).
With the Ansatz $\mathbf{n}=\{0,0,\sin f(x),\cos f(x)\}$ 
we have  
\beq
\mathcal{L} = -\frac{1}{2}(\p_x f)^2 - \frac{1}{2}M^2\sin^2 f.
\eeq
This Lagrangian density is the sine-Gordon model admitting a
domain-wall solution 
\beq
f = 2\tan^{-1}\exp(\pm M x). \label{eq:SG}
\eeq
The domain wall in this type of model was first studied in Ref.~\cite{Kudryavtsev:1999zm}.
The most general solution is given by 
\beq
\mathbf{n} = \{b_1\sin f(x),b_2\sin f(x),b_3\sin f(x),\cos f(x)\},
\eeq
which has moduli in the form of a constant three-vector $\mathbf{b}$
with unit length $\mathbf{b}^2=1$.
These are Nambu-Goldstone (NG) modes due to the spontaneously broken
$O(3)$ symmetry, which is broken down to $O(2)$ in the presence of the 
domain wall (\ref{eq:SG}).
These $S^2$ moduli of the domain wall were discussed in 
Refs.~\cite{Losev:2000mm,Nitta:2012wi,Nitta:2012rq}.

%%%%%%%%%%%%%
\subsection{Model 2: the vortex}
\label{sec:model2}

\begin{figure}[!tp]
\begin{center}
\mbox{\subfigure[]{\includegraphics[width=0.48\linewidth]{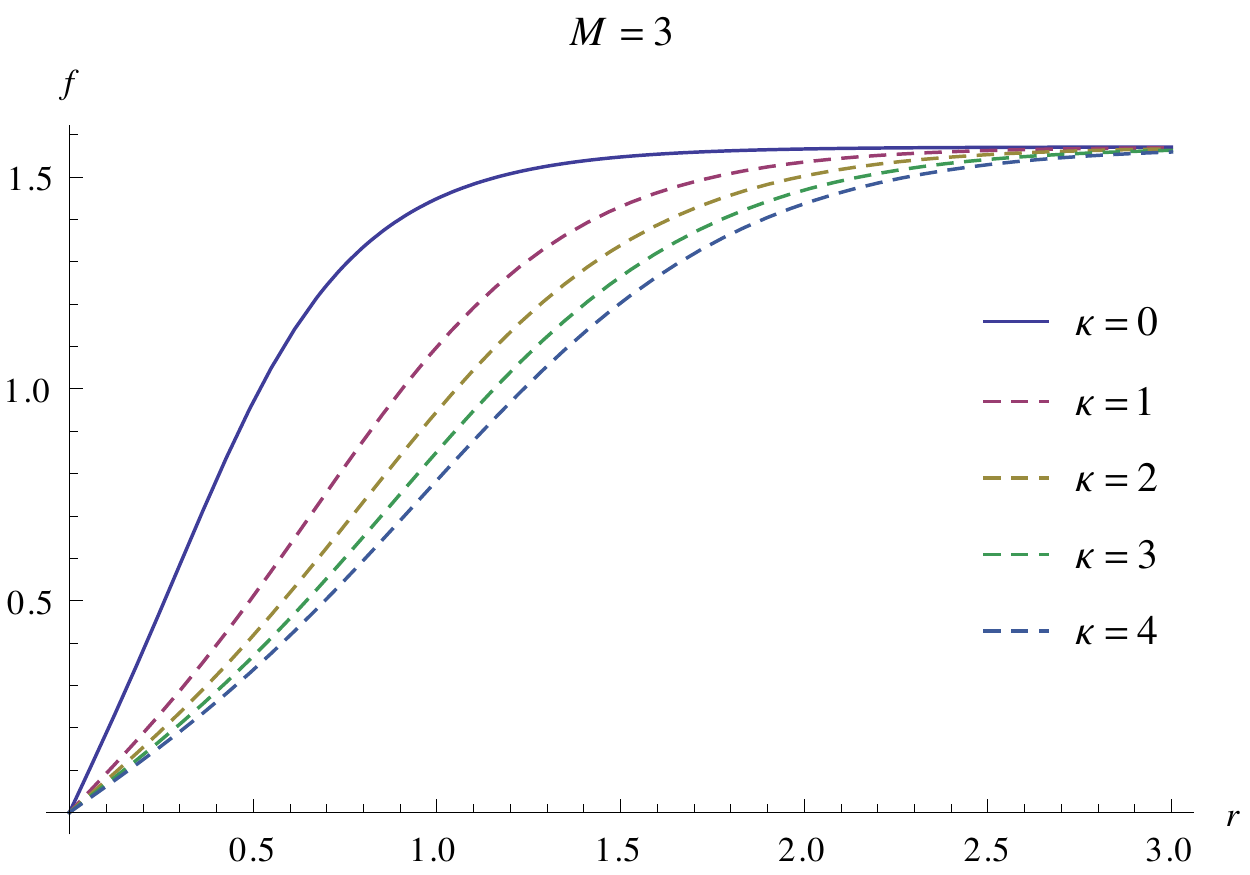}}\quad
\subfigure[]{\includegraphics[width=0.48\linewidth]{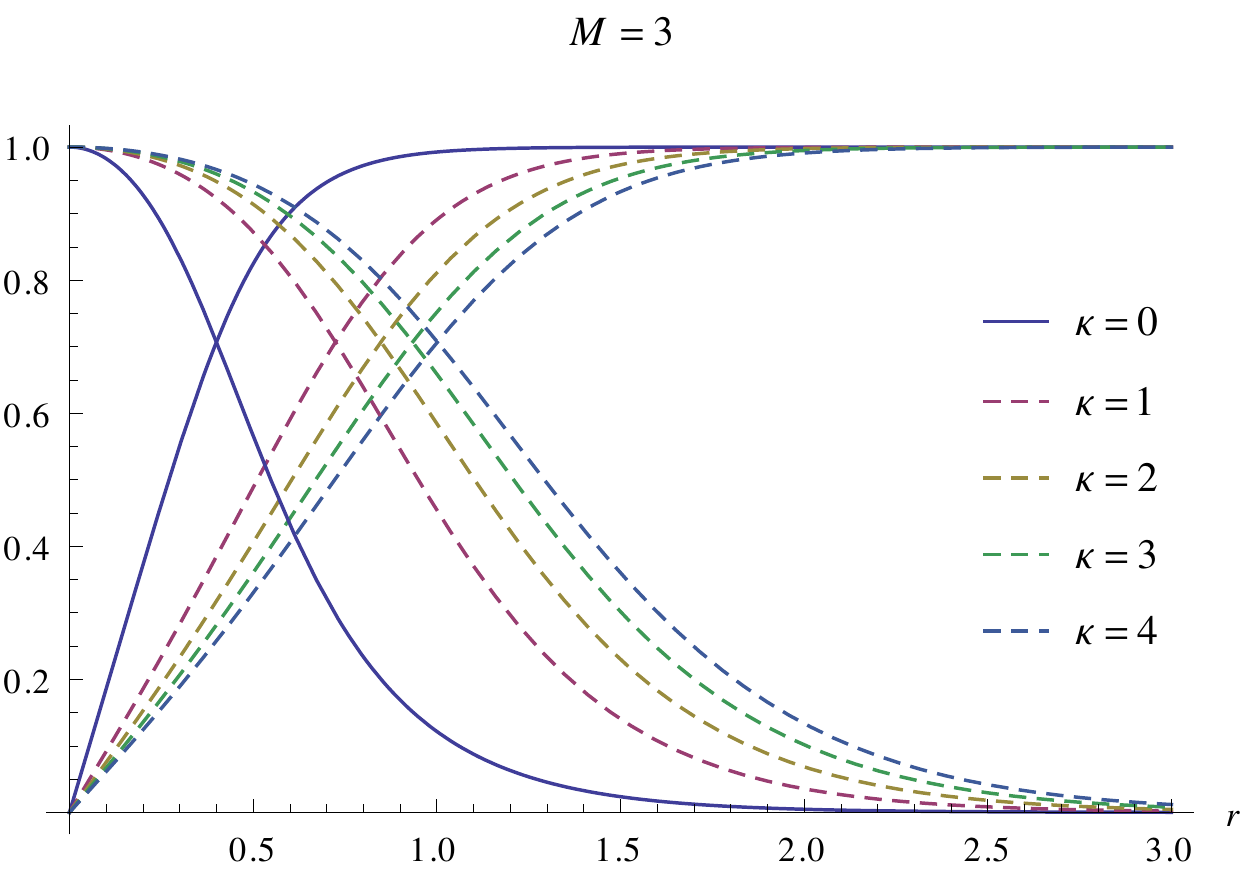}}}\\
\mbox{\subfigure[]{\includegraphics[width=0.48\linewidth]{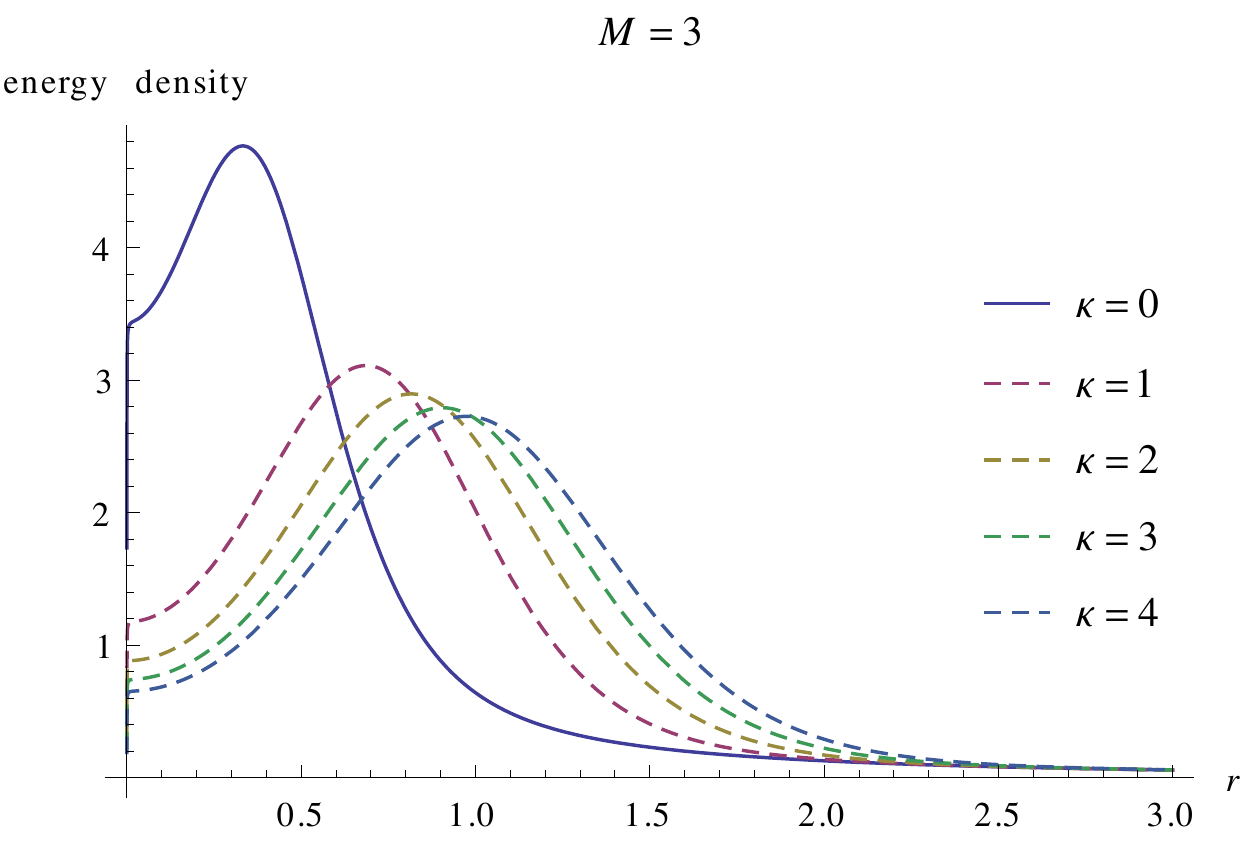}}}
\caption{(a) Vortex profile function $f$ without the Skyrme term
  $\kappa=0$ (blue solid curve) and with the Skyrme term
  $\kappa=1,\ldots,4$ (dotted curves) for mass $M=3$.
  (b) Condensate fields $|\phi_1|=\sin f$ and $|\phi_2|=\cos f$
  ($\phi_1$ vanishes at the origin).
  (c) Corresponding energy densities. 
\label{fig:vortex}}
\end{center}
\end{figure}

Model 2 allows for global vortices.
The vortex of $\phi_1$ traps $\phi_2$ in its core and carries a $U(1)$
modulus being the phase of $\phi_2$. 
For constructing the vortex, we use the following Ansatz
\beq
\phi_1 = \sin f(r) e^{i\phi}, \qquad
\phi_2 = \cos f(r),
\eeq
where $r,\phi$ are polar coordinates in the plane. This simplifies the 
Lagrangian density to
\beq
-\mathcal{L} = 
\frac{1}{2}f_r^2
+\frac{1}{2r^2}\sin^2 f
+\frac{\kappa}{2r^2}\sin^2(f) f_r^2
+\frac{1}{8} M^2 \sin^2(2f),
\eeq
and the equation of motion reads
\beq
f_{rr} + \frac{1}{r} f_r -\frac{1}{2r^2}\sin 2f
+\frac{\kappa}{r^2}\sin^2 f\left(f_{rr} - \frac{1}{r} f_r\right)
+\frac{\kappa}{2r^2}\sin(2f) f_r^2
-\frac{1}{4} M^2 \sin 4f = 0.
\eeq
The boundary conditions for the vortex system are 
\beq
f(0) = 0, \qquad
f(\infty) = \frac{\pi}{2}.
\eeq
Numerical solutions are shown in Fig.~\ref{fig:vortex} for
$\kappa=0,1,\ldots,4$. 
%%%%%%%%%%%%%%%%%%%%%%
%%%%%%%%%%%%%%%%%%%%%%
Notice that when $\kappa$ is turned on, the vortex widens up and the
energy density at the origin drops significantly. 

This vortex is a particular solution, while the most general solution
is given by
\beq
\phi_1 = \sin f(r) e^{i\phi}, \qquad
\phi_2 = \cos f(r) e^{i\zeta},
\eeq
where $\mathbf{b}=\{\cos\zeta,\sin\zeta\}$ is a constant $U(1)$
modulus.

%%%%%%%%%%%%%%%%%%%%%%%%%%%%%%%%%%
\section{Effective theory approach}\label{sec:effective}

In this section we will review the results of 
Ref.~\cite{Gudnason:2014gla}. The main idea is to take a soliton solution
and integrate out the soliton to obtain the effective theory for the
moduli inhabiting the soliton in question. Here we will consider only
the leading-order effective Lagrangian. For details and a discussion
of the expansion, see Ref.~\cite{Gudnason:2014gla}. We will next consider
each model in turn. 

\subsection{Model 1: Skyrmions as  baby Skyrmions inside 
a domain wall}\label{sec:eff-wall-skyrmion}

We start with the Lagrangian density \eqref{eq:masterL_dimless} and
integrate over the codimension of the domain wall. 
Here we will take just the leading-order effective Lagrangian and
neglect the backreaction of the baby soliton to the domain wall. This
is a good approximation when there is a separation of scales between
the domain wall mass and the baby-soliton's typical scales. 
At leading-order the domain wall is flat which leads to a drastic
simplification and the integration over its codimension yields
\begin{align}
-\mathcal{L}^{\rm eff} &= 
  \left(\frac{a_{2,0}}{M} + \kappa a_{2,2} M\right) \;
  (\p_\alpha\mathbf{b})^2 
  +\left(\frac{\kappa a_{4,0}}{2M} + c_6 a_{4,2} M\right) 
  \big(\p_\alpha\mathbf{b}\times\p_\beta\mathbf{b}\big)^2
- \frac{a_{2,0} m_3^2}{M} b_3^{a_3},
\label{eq:Lcodim1}
\end{align}
where $M$ is the mass scale of the domain wall, $\alpha,\beta=t,y,z$,
$a_3=1,2$ and we have defined dimensionless constants as follows 
\begin{align}
a_{k,\ell} &\equiv 
\frac{M}{2} \int dx \; \sin^{k} f \left(\frac{\p_x f}{M}\right)^{\ell} 
= \frac{1}{2}\int d\xi \sech^{k+\ell}\xi 
= \frac{\sqrt{\pi}\,\Gamma\!\left(\frac{k+\ell}{2}\right)}
  {\Gamma\!\left(\frac{1+k+\ell}{2}\right)},
\end{align}
where $\xi=M x$ and the last two equalities have been evaluated using
the flat domain wall \eqref{eq:SG} and the result is given in terms of
the gamma function. Notice that the coefficients depend only on the
sum of $k$ and $\ell$ and so we can evaluate them as 
$a_{k,\ell} = a_{k+\ell}$: 
\beq
a_{2} = 1, \quad
a_{4} = \frac{2}{3}, \quad
a_{6} = \frac{8}{15}, \quad
\cdots
\eeq
Inserting the coefficients we get
\begin{align}
-\mathcal{L}^{\rm eff} &= 
  \left(\frac{1}{M} + \frac{2}{3}\kappa M\right) \;
  (\p_\alpha\mathbf{b})^2 
  +\left(\frac{\kappa}{3M} + \frac{8}{15} c_6 M\right) 
  \big(\p_\alpha\mathbf{b}\times\p_\beta\mathbf{b}\big)^2
- \frac{m_3^2}{M} b_3^{a_3}.
\label{eq:Lcodim1_coeff}
\end{align}

The existence of a baby-Skyrmion living on the domain wall requires a
non-vanishing $\kappa>0$ or $c_6>0$ as well as a non-zero $m_3$. The
size of the baby-Skyrmion can be estimated by a scaling argument
\cite{Derrick:1964ww} to be
\beq
\frac{1}{L} \sim 
\sqrt[4]{\frac{15m_3^2}{5\kappa + 8c_6 M^2}}. 
\eeq

A second kind of soliton which can inhabit the domain wall is the
lump, which exists when $\kappa=c_6=m_3=0$ and it will possess a size
modulus \cite{Nitta:2012wi}. 

The (full 3D) baryon charge is composed by the domain wall charge and
the baby-Skyrmion charge and is given by 
\begin{align}
B = \frac{1}{\pi} \int d^3x \; \mathcal{Q} f_x = Q,
\end{align}
where we have used that there is only a single domain wall in this
setup while
\beq
\mathcal{Q} = \frac{1}{8\pi} \epsilon^{ij} 
  \mathbf{n}\cdot\del_i\mathbf{n}\times\del_j\mathbf{n},
\eeq
is the baby-Skyrmion charge density and $Q$ the baby-Skyrmion number
(charge).

%%%%%%%%%%%%%%%%%%%%%%%%%%%%%%%%
\subsection{Model 2: Skyrmions as kinks on a vortex}
\label{sec:eff-vortex-skyrmion}

The next and final type of soliton we will consider is the vortex
which has codimension two and a single world-volume direction.
We again take the Lagrangian density \eqref{eq:masterL_dimless} and 
integrate over the two codimensions of the vortex to obtain 
\begin{align}
-\mathcal{L}^{\rm eff} = \left[\frac{a_{2,0,0}}{M^2}
    + \kappa(a_{2,2,0} + a_{2,0,2}) 
    + 2c_6 a_{2,2,2} M^2\right] (\p_\alpha \mathbf{b})^2 
  - \frac{a_{2,0,0} m_3^2}{M^2} b_1^2,
\end{align}
where $M$ is the mass scale of the vortex, $\alpha=t,z$ and the
dimensionless coefficients in the effective Lagrangian density read
\begin{align}
a_{k,\ell,m}
\equiv \pi M^{2-\ell-m} \int dr \; r^{1-\ell} \cos^k f\sin^\ell f (f_r)^m.
\label{eq:vortex_coefficients}
\end{align}
This effective theory possesses sine-Gordon kinks. Hence, the vortex
can bear sine-Gordon kinks in terms of its twisted $S^1$ modulus and
each of these kinks correspond to Skyrmions in the full 3-dimensional
theory. 

Unfortunately, the vortex is not analytically integrable and hence we
need to evaluate the coefficients numerically. As we have defined the
coefficients in a dimensionless manner, they do not depend on the
value of the vortex mass scale, $M$, but they do depend on the value
of the fourth-order derivative term, $\kappa$ (or rather the
combination $\kappa M^2$). We give a set of numerically evaluated
coefficients for the effective Lagrangian density in
Tab.~\ref{tab:a_vortex_numerical}. 

\begin{table}[!btp]
\begin{center}
\caption{Coefficients for the effective Lagrangian density for
  sine-Gordon kinks living on a straight vortex for various values of
  $\kappa M^2$. }
\label{tab:a_vortex_numerical}
\begin{tabular}{c||ccccc}
$\kappa M^2$ & $0$ & $1$ & $2$ & $3$ & $4$\\
\hline\hline
$a_{2,0,0}$ & 0.5106  & 0.7224  & 0.8678  & 0.9866  & 1.090\\
$a_{2,2,0}$ & 0.1616  & 0.1550  & 0.1519  & 0.1499  & 0.1484\\
$a_{2,0,2}$ & 0.1745  & 0.1816  & 0.1852  & 0.1877  & 0.1896\\
$a_{2,2,2}$ & 0.06072 & 0.04172 & 0.03438 & 0.03007 & 0.02712
\end{tabular}
\end{center}
\end{table}

Using again a scaling argument \cite{Derrick:1964ww}, we can estimate
the size of the sine-Gordon kink
\beq
\frac{1}{L} \sim \sqrt{\frac{a_{2,0,0}m_3^2}
{a_{2,0,0} + \kappa(a_{2,2,0} + a_{2,0,2}) M^2 + 2c_6 a_{2,2,2} M^4}},
\label{eq:kinksizeestimate}
\eeq
where the coefficients $a$ are functions of $\kappa M^2$, as shown in 
Tab.~\ref{tab:a_vortex_numerical}. As examples, we can calculate the
kink sizes, see Tab.~\ref{tab:kink_sizes}.
\begin{table}[!tbp]
\begin{center}
\caption{Rough size estimate of sine-Gordon kinks using
  Eq.~\eqref{eq:kinksizeestimate} as function of the parameters of the
  effective theory. }
\label{tab:kink_sizes}
\begin{tabular}{c||rrr}
\raisebox{-2pt}{$\kappa M^2$}${\bf \backslash}$\raisebox{2pt}{$c_6 M^4$} 
  & $0$ & $1$ & $81$\\
\hline\hline
$0$ & $m_3$ & $0.90m_3$ & $0.22m_3$ \\
$1$ & $0.83m_3$ & $0.80m_3$ & $0.30m_3$\\
$9$ & $0.57m_3$ & $0.57m_3$ & $0.44m_3$
\end{tabular}
\end{center}
\end{table}
Thus from these rough estimates, we learn that the sixth-order
derivative term induces coefficients in the effective Lagrangian for
the kink which increases its size (for fixed masses). The fourth-order
derivative term also leads to an increase in the kink size, but to a
lesser extent.

The (full 3D) baryon charge is composed by the vortex charge and the
kink charges and is given by 
\begin{align}
B = \frac{1}{16\pi^2} \int d^3x \; \frac{1}{r}\sin(f) f_r \zeta_z 
= Q [\zeta]^{z=z_2}_{z=z_1} = Q P,
\end{align}
where $Q$ is the winding number of the vortex and $P$ is the number
kinks on the string.

%%%%%%%%%%%%%%%%%%%%%%%%%%%%%%
\section{Numerical solutions}\label{sec:numerical}

In this section, we provide explicit numerical solutions.
Solutions of the baby-Skyrmion type in model 1 were already obtained
in Ref.~\cite{Gudnason:2014nba}. 
Here we will add a new lump solution living on the domain wall, which
also carries baryon charge.
Our other new findings are confined Skyrmions residing on the vortex
string in model 2. 
For both cases, we need the deformation $V_2$ of the potentials 
for flat host solitons, while we do not need it for solutions
possessing an $S^n$ world-volume. 
On the contrary, we need no higher-derivative terms for flat host
solitons, while we do need them for the solutions possessing an $S^n$ 
world-volume for the stability. An exception to the rule is the
baby-Skyrmion living on the domain wall, which needs both a
higher-derivative term as well as the potential $V_2$ (but the lump on 
the domain wall needs neither of these). 

For both models we use the relaxation method on a cubic square-lattice
of size $81^3$ (lattice points). We fix the boundary conditions
corresponding to the host solitons as described in
Sec.~\ref{sec:wall-vortex} and choose appropriate initial conditions
for the baby-solitons in question. Then we relax the initial guess
until the solution to the equations of motion is obtained with the
required precision. A cross check of the solutions is the calculation
of the topological baryon charge. We will now take the two models in
turn.

%%%%%%%%%%%%%%%%%%%%%%%%%%%
\subsection{Model 1: Skyrmions trapped inside a domain wall}
\label{sec:wall-skyrmion}

\begin{figure}[!pt]
\begin{center}
\mbox{\subfigure[\ isosurfaces]{
\includegraphics[width=0.3\linewidth]{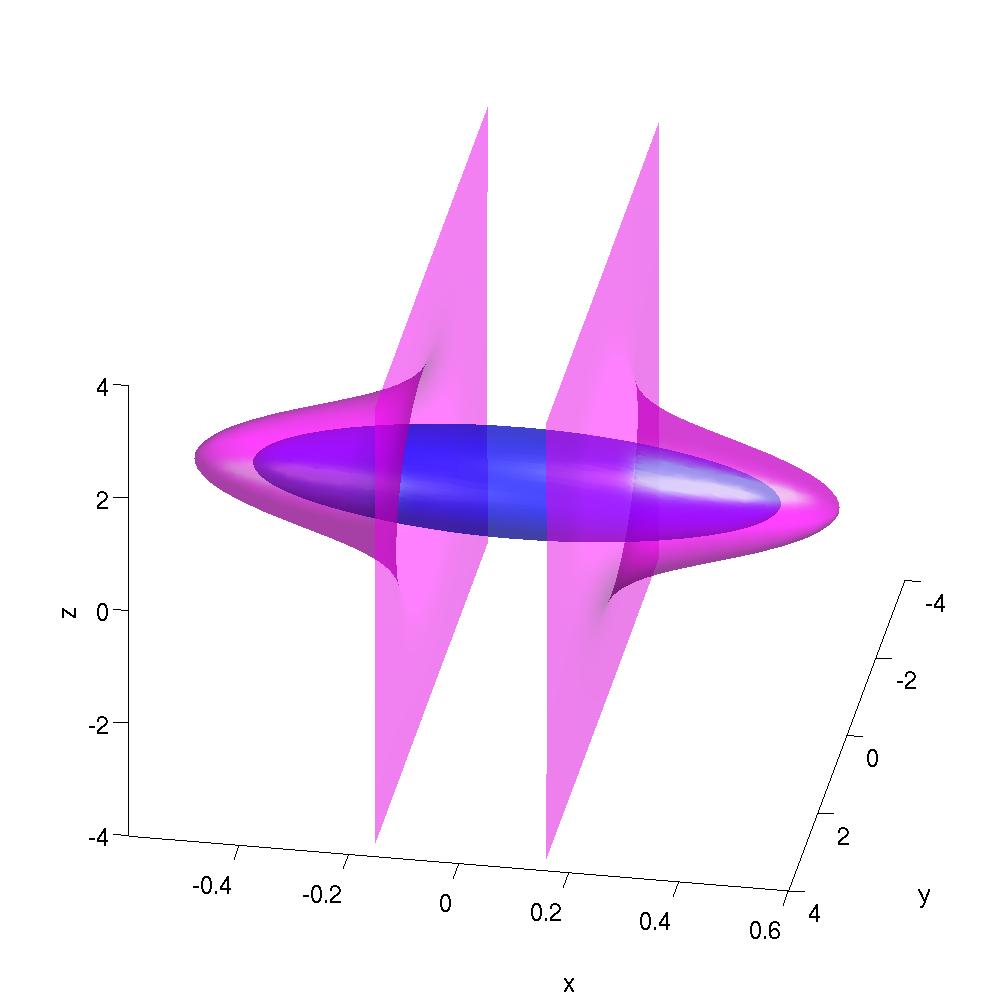}}
\subfigure[\ energy density]{
\includegraphics[width=0.3\linewidth]{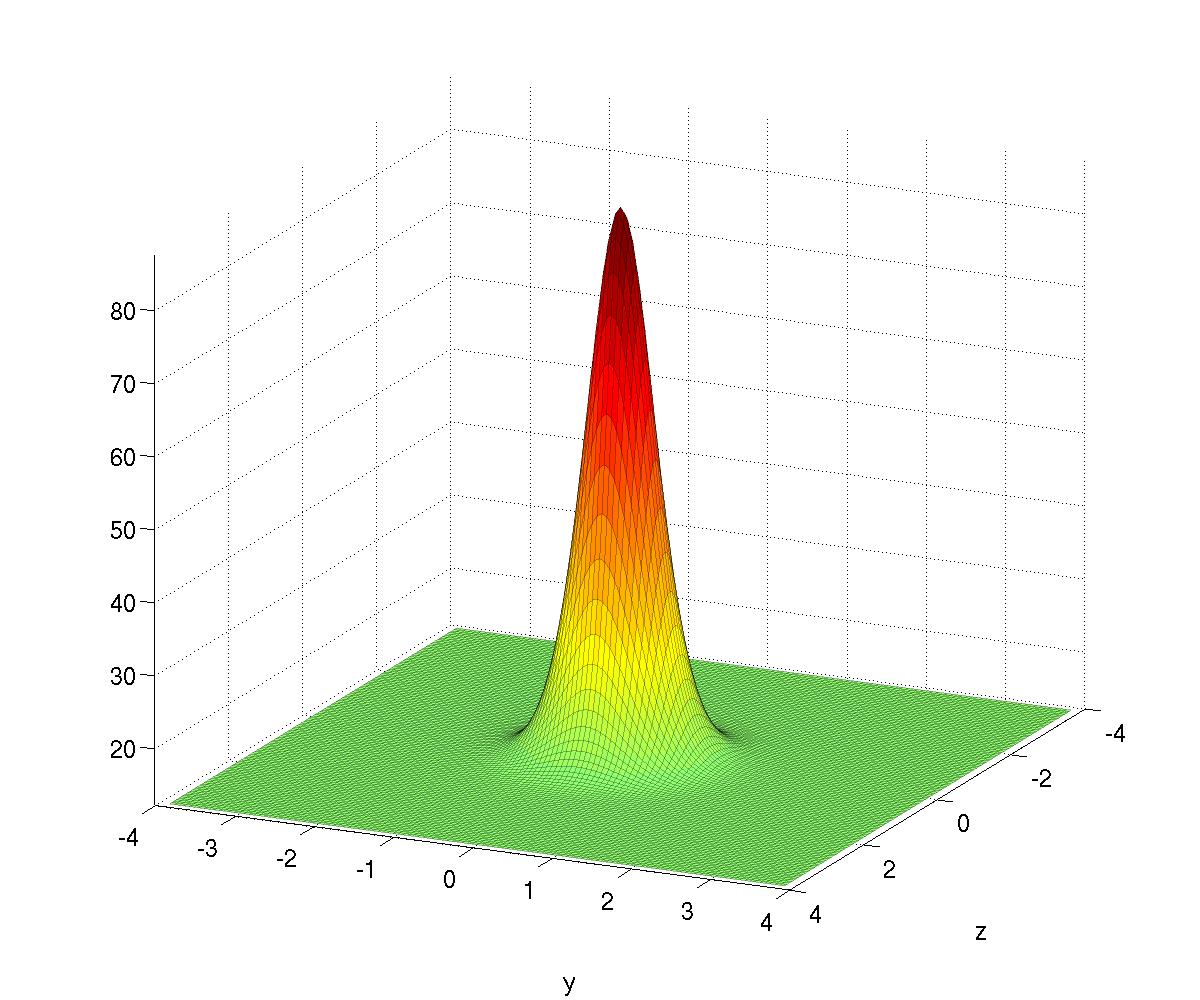}}
\subfigure[\ baryon charge density]{
\includegraphics[width=0.3\linewidth]{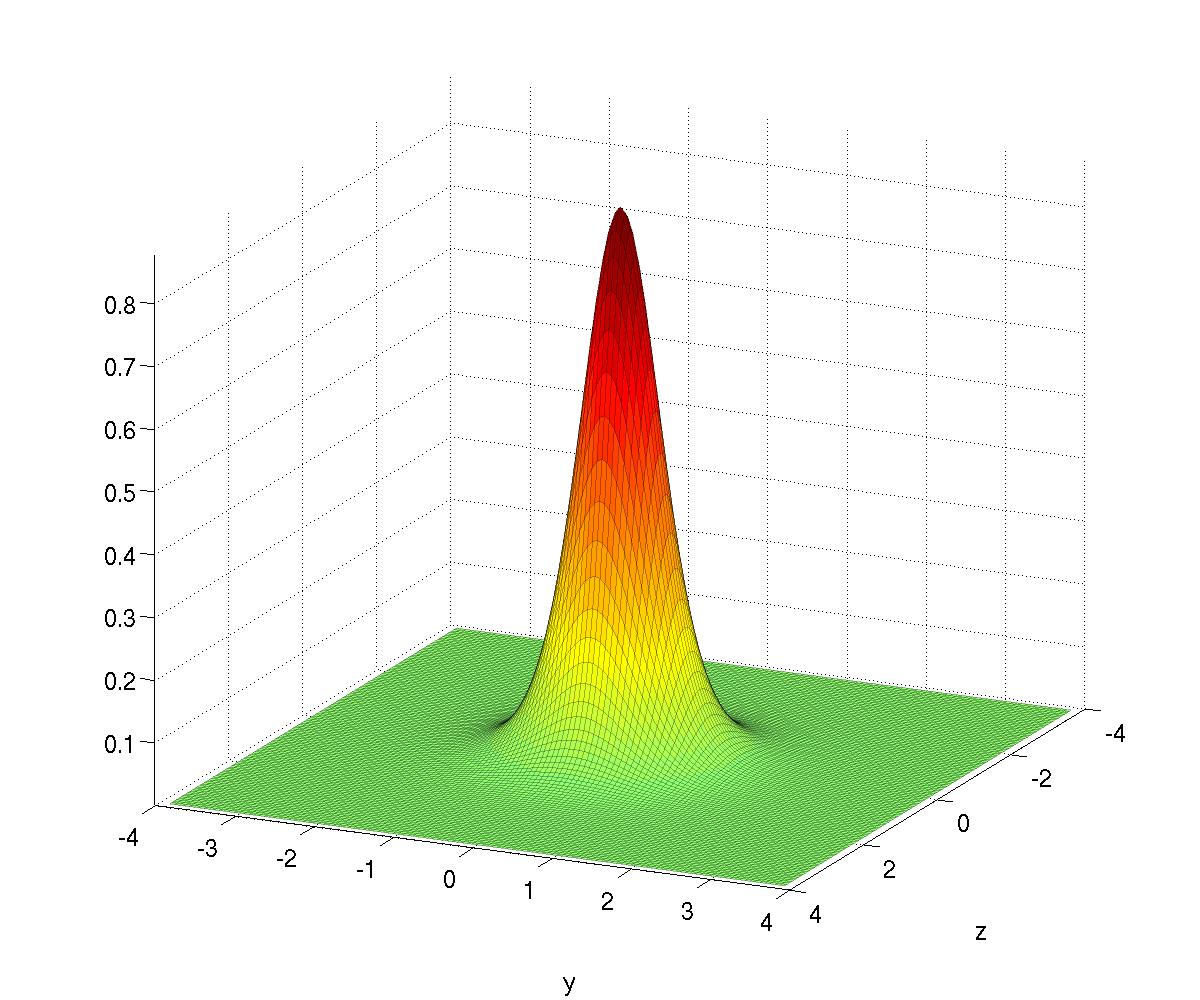}
}}
\caption{The baby-Skyrmion living on the domain wall: 
  (a) 3D view of isosurfaces for the domain wall on which a
  baby-Skyrmion resides; the magenta surfaces represent the energy
  isosurfaces at a third of the maximum of the energy and 
  the blue surface at the center shows the baryon charge isosurface,
  at half its maximum value. (b) and (c) show respectively the energy
  density and baryon charge density at a $yz$-slice in the middle of
  the domain wall (at $x=0$). The calculation is done on a $129^3$
  cubic lattice, $B^{\rm numerical}=0.998$ and the potential used is
  $V_2$ with $a_3=1$ and $M=4,m_3=2$. 
  This figure is taken from Ref.~\cite{Gudnason:2014nba}. }
\label{fig:DW_SL}
\end{center}
\end{figure}

We begin with the (exact) domain wall solution of
Sec.~\ref{sec:model1} and add two types of baby-solitons. The first
example is the baby-Skyrmion, which was obtained in
\cite{Gudnason:2014nba} and we will only review it here for
completeness. This solution needs both the potential $V_2$ as well as
a higher-order derivative term. In Fig.~\ref{fig:DW_SL} we show the
baby-Skyrmion with $V_2$ setting $a_3=1$ and $\kappa=1,c_6=0$ (thus
only the Skyrme term is active), which is taken from
\cite{Gudnason:2014nba}. 

\begin{figure}[!hpt]
\begin{center}
\mbox{\subfigure[\ isosurfaces]{
\includegraphics[width=0.3\linewidth]{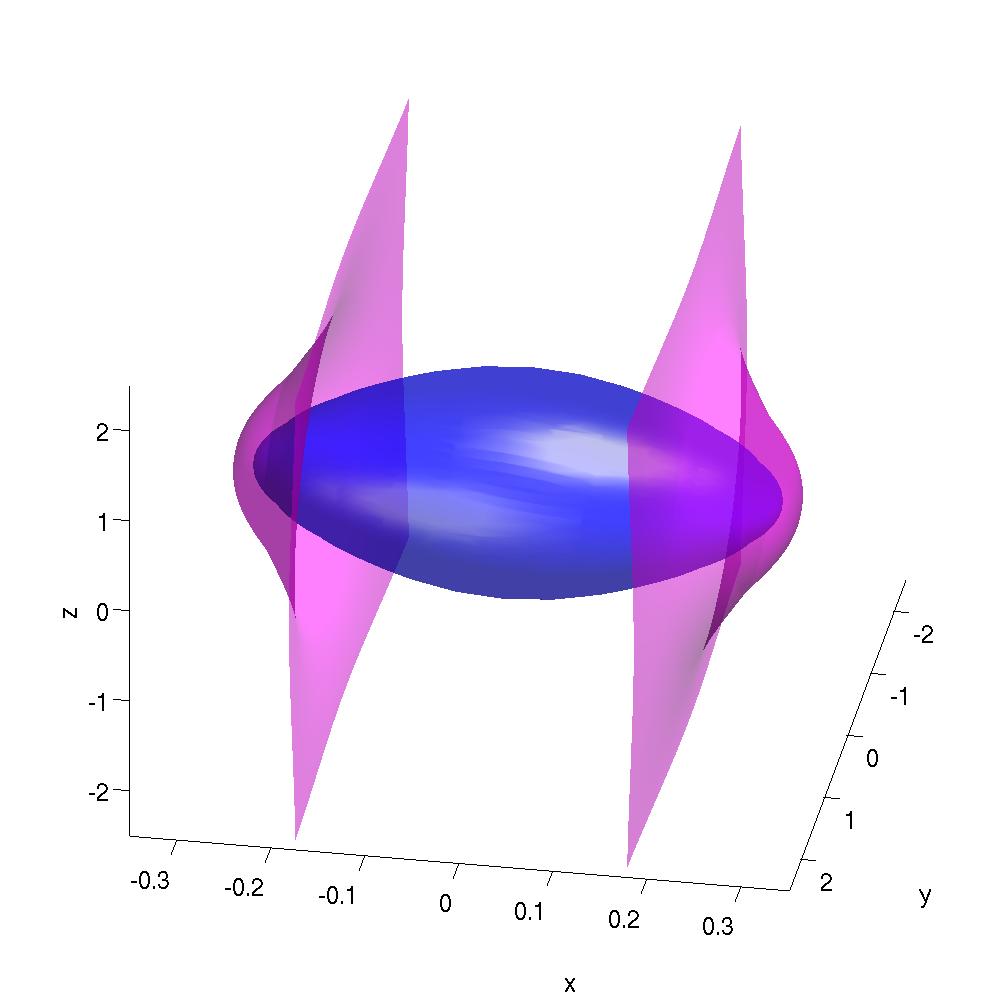}}
\subfigure[\ energy density]{
\includegraphics[width=0.3\linewidth]{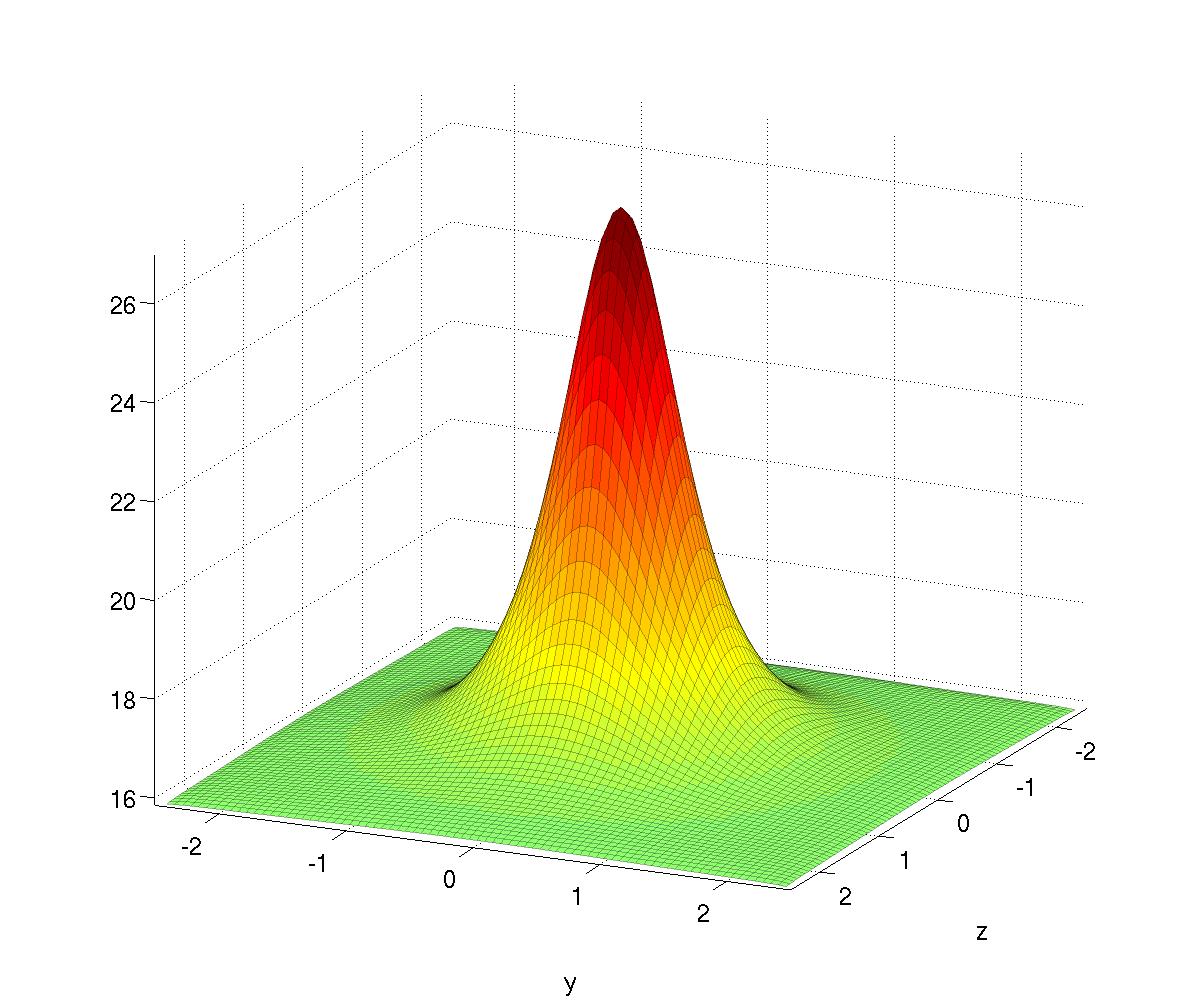}}
\subfigure[\ baryon charge density]{
\includegraphics[width=0.3\linewidth]{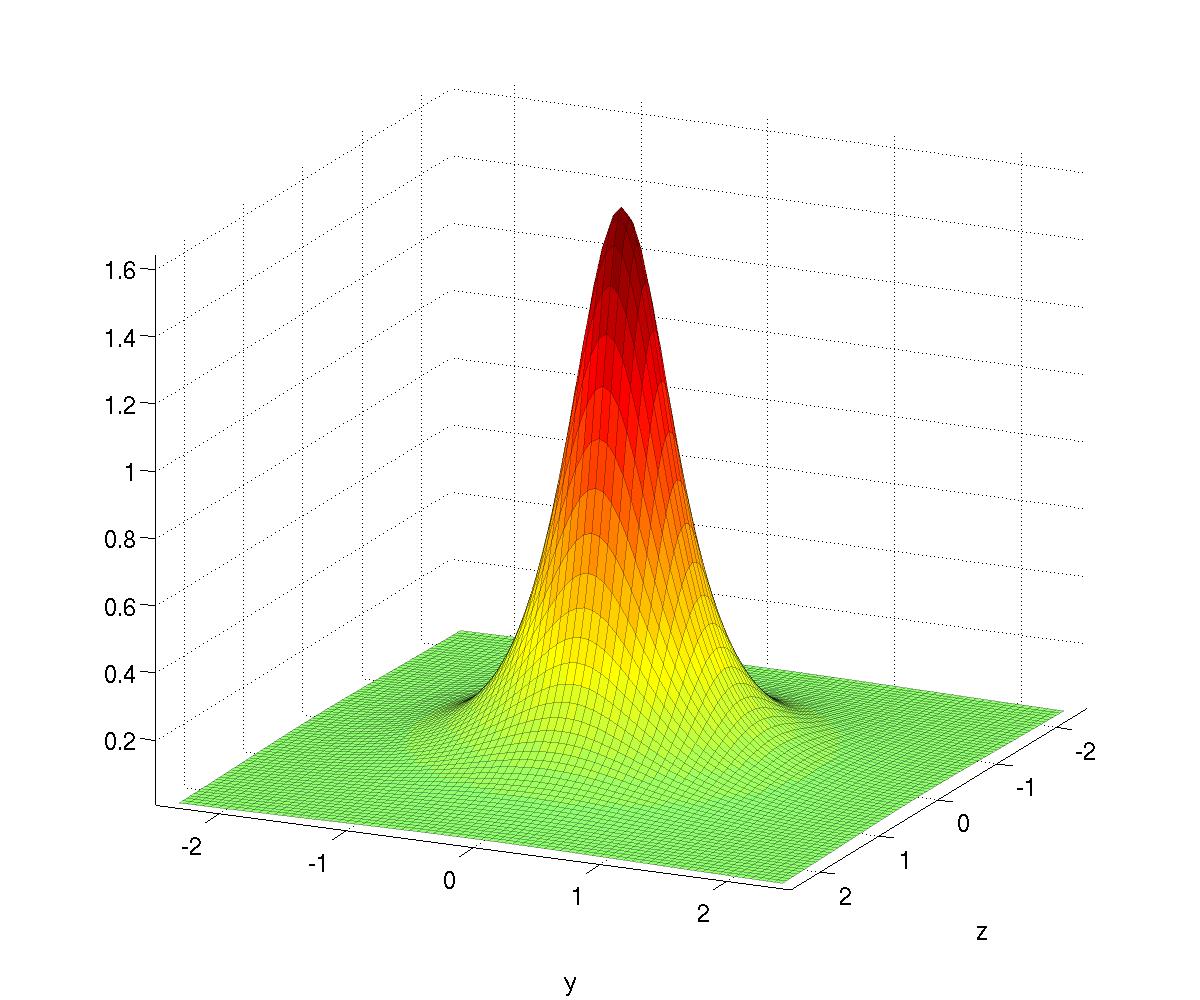}}}
\caption{The domain wall with a trapped Skyrmion 
  in the theory without higher-derivative and potential terms. 
  (a) 3D view of isosurfaces for the energy density and baryon charge
  density as in Fig.~\ref{fig:DW_SL}. (b) and (c) show respectively
  the energy density and baryon charge density at a $yz$-slice in the
  middle of the domain wall (at $x=0$). The calculation is done on an
  $81^3$ cubic lattice, with $M=4$ and $B^{\rm numerical}=0.877$. }
\label{fig:DW_LUMP}
\end{center}
\end{figure}

The next solution, which is new, is the domain wall with a lump
solution inside. This solution also carries a full unit of
3-dimensional baryon (Skyrme) charge. It is obtained for $V_2=0$ and
no higher-derivative terms, i.e.~$\kappa=0,c_6=0$. The numerical
solution is shown in Fig.~\ref{fig:DW_LUMP}. Notice that the lump has
a size modulus and can thus take on any size. We also do not capture
the full baryon charge because we resolve only the center of the lump
with the finite lattice points. This is not a problem of the solution
but of the lattice size. A larger lattice will capture more of the
baryon charge (or alternatively a small lump on the same lattice with
a worse resolution). 

%%%%%%%%%%%%%%%

%%%%%%%%%%%%%%%%%%%%%%%%%%%
\subsection{Model 2: Skyrmions confined by vortices}
\label{sec:vortex-skyrmion}

In this section we take the vortex solution of Sec.~\ref{sec:model2}
and add a sine-Gordon kink on its world-volume. 
As already mentioned, for the straight vortex we need a finite
potential $V_2$ for the baby-soliton. If we choose a linear potential,
i.e.~$a_3=1$, the kink on the vortex corresponds to a full unit of
baryon charge, while for the quadratic potential, i.e.~$a_3=2$, each
kink corresponds to half a unit of baryon charge. 

The straight vortex possesses sine-Gordon kinks if the potential $V_2$ 
is turned on even without higher-derivative terms. 
The presence of the Skyrme term (the fourth-order derivative term)
widens both the vortex itself and the kink living on the vortex (see
Sec.~\ref{sec:eff-vortex-skyrmion}, whereas the sixth-order derivative
term does not alter the vortex solution, but it does widen the kink --
even more than the Skyrme term, see Tab.~\ref{tab:kink_sizes}. 

\begin{figure}[!tbh]
\begin{center}
\mbox{\subfigure[\ isosurfaces]{
\includegraphics[width=0.4\linewidth]{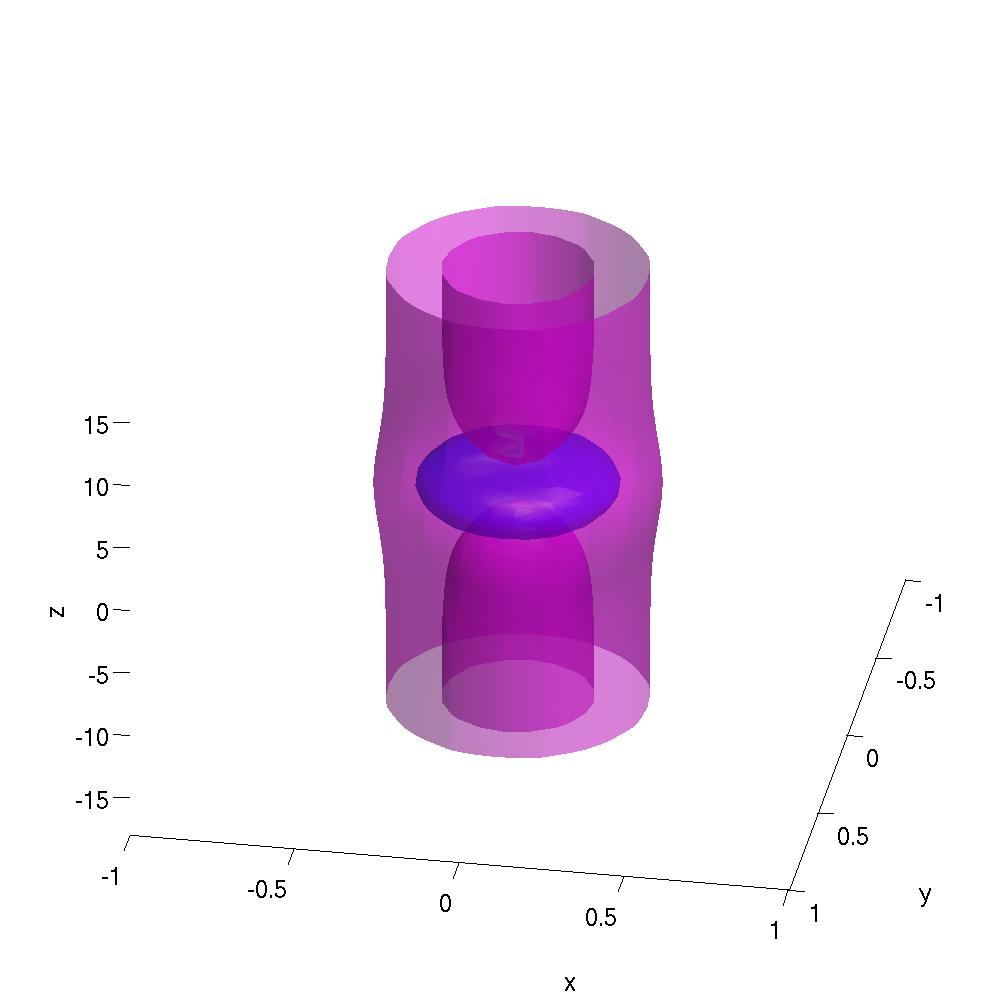}}
\subfigure[\ energy density]{
\includegraphics[width=0.4\linewidth]{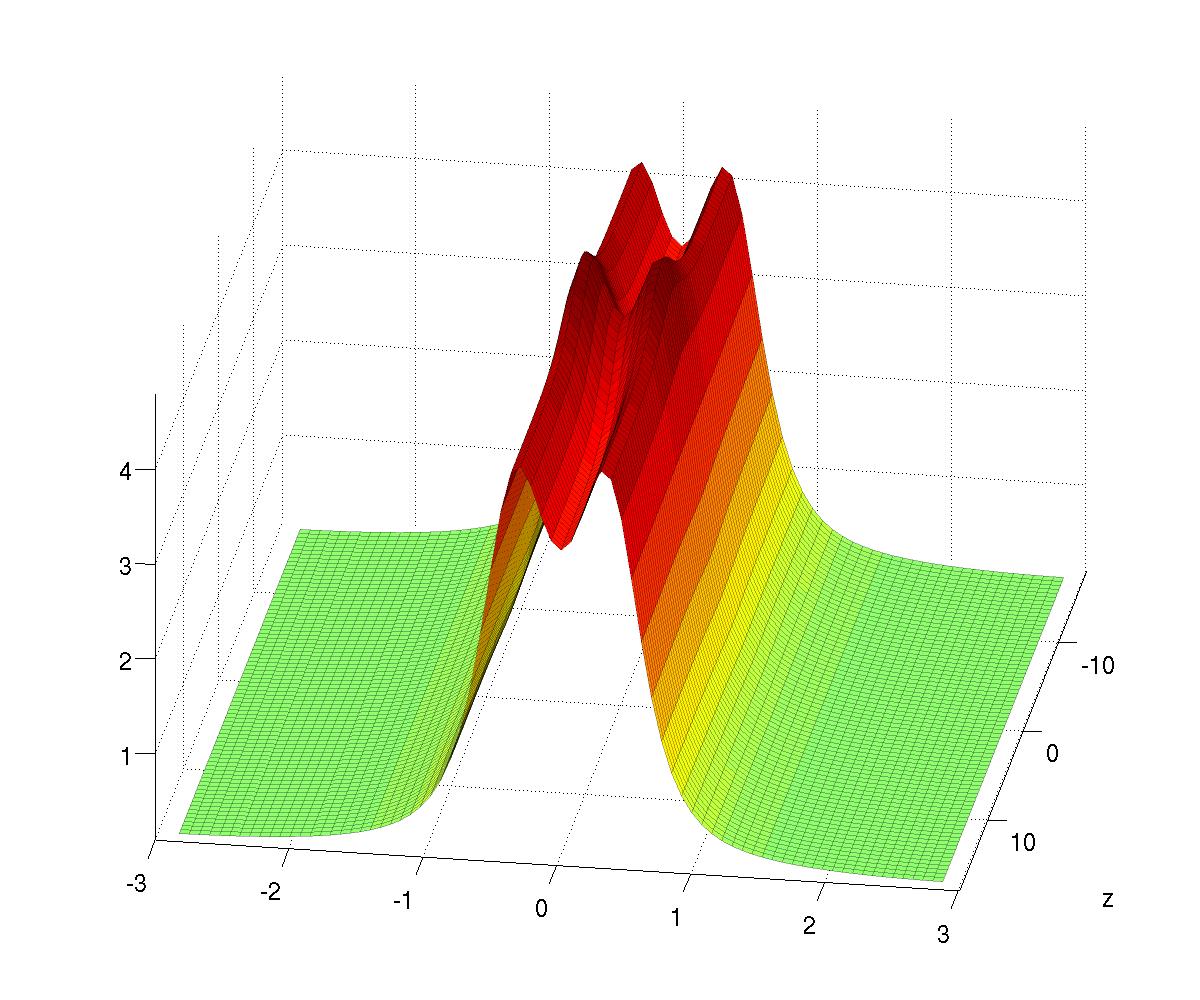}}}
\mbox{\subfigure[\ baryon charge density]{
\includegraphics[width=0.4\linewidth]{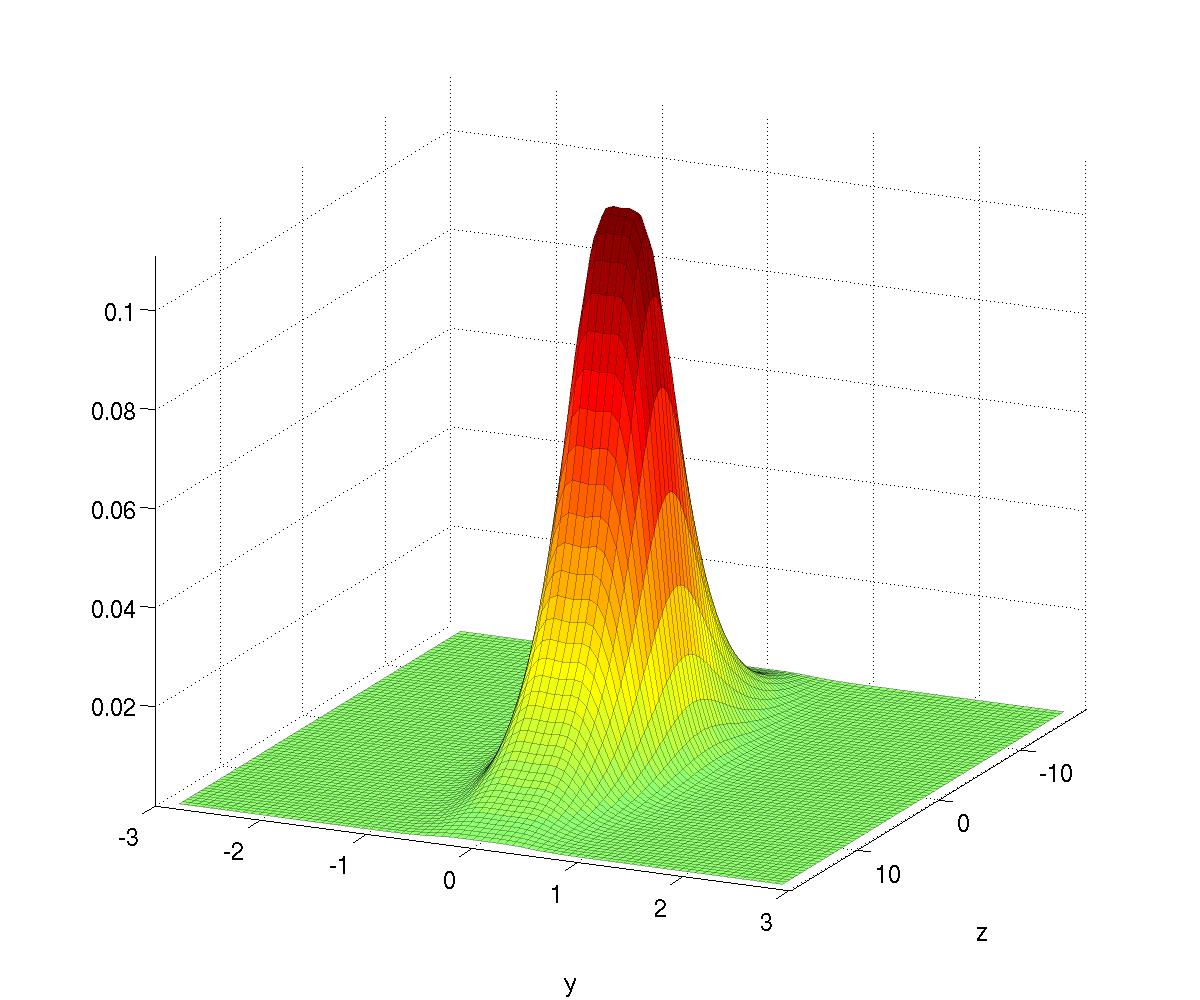}}
\subfigure[\ kink energy density]{
\includegraphics[width=0.4\linewidth]{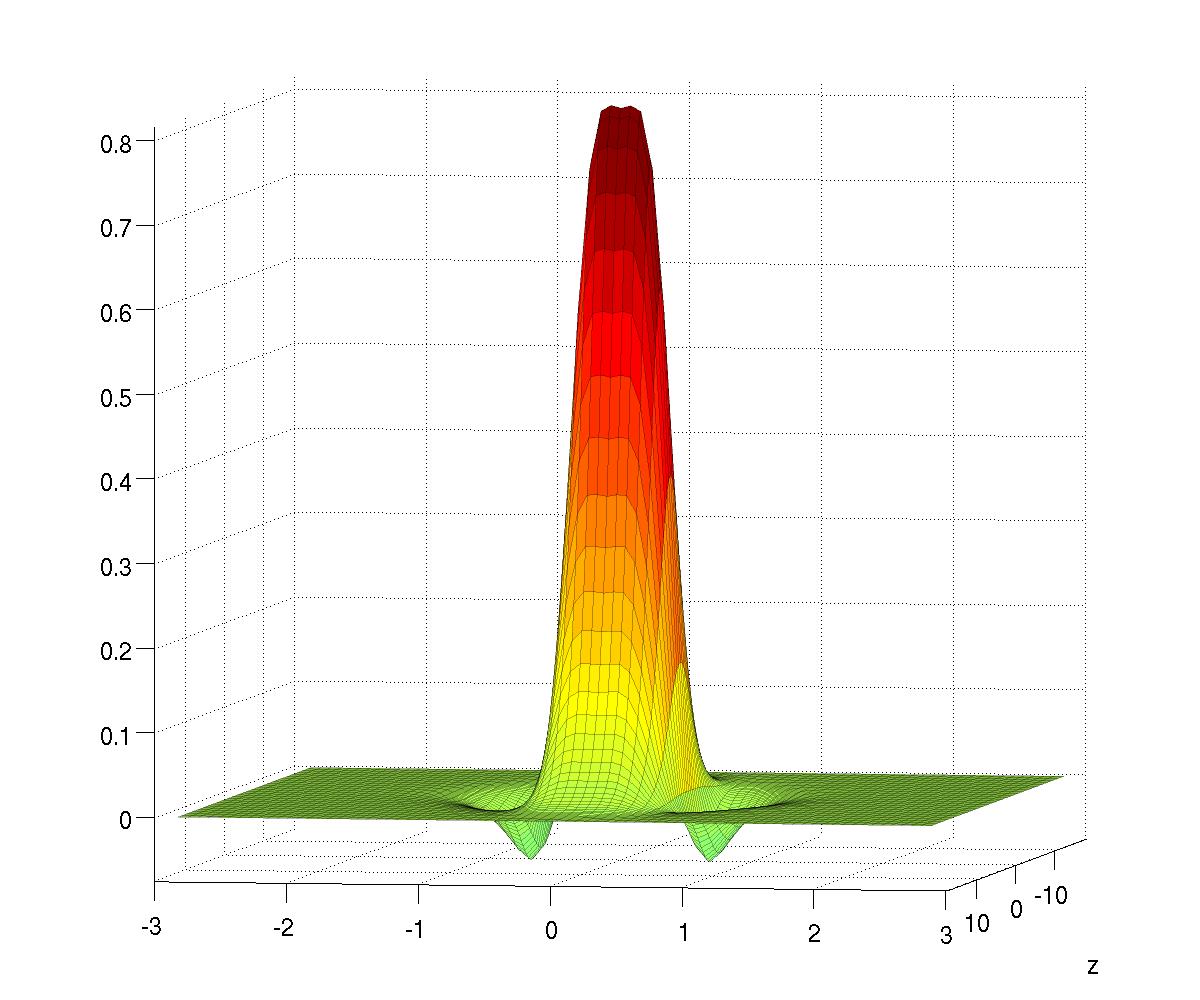}}}
\caption{The vortex with a trapped Skyrmion which is manifested as a
  sine-Gordon kink on its world-volume, in the theory with no
  higher-derivative terms. 
  (a) 3D view of isosurfaces for the energy density and baryon charge
  density. (b) and (c) show respectively
  the energy density and baryon charge density at a $yz$-slice through
  the vortex (at $x=0$). (d) shows the energy of the kink (which is
  the total energy with the vortex energy subtracted off). Notice the
  negative dips on each side of the peak in the kink energy; we
  interpret those as binding energy. 
  The calculation is done on an $81^3$ cubic lattice, with $M=3$,
  $m_3=0.22$ and the baryon charge is $B^{\rm numerical}=0.991$. }
\label{fig:vortexkink_ko}
\end{center}
\end{figure}

First we will present the vortex with a full sine-Gordon kink living 
on its world-volume in the case of no higher-derivative terms, see
Fig.~\ref{fig:vortexkink_ko}. This kink was made with kink mass
$m_3=0.22$, vortex mass $M=3$ and the kink length was measured as
\beq
L_{\rm kink} = \sqrt{\frac{\int d^3x \; z^2
    \mathcal{E}_{\rm kink}}{\int d^3x \; \mathcal{E}_{\rm kink}}} 
  \sim 3.35, 
\label{eq:kinklengthmeasure}
\eeq
which one can compare with the analytic formula
$\frac{\pi}{2\sqrt{3}m_3}\simeq 4.14$. The reason for the smaller
value in the numerical result is due to the binding energy present on
both sides of the center of the kink, see
Fig.~\ref{fig:vortexkink_ko}d. 

%%%%%%%%%%%%%%%%%%%%%%
%%%%%%%%%%%%%%%%%%%%%%

\begin{figure}[ht]
\begin{center}
\mbox{\subfigure[\ isosurfaces]{
\includegraphics[width=0.4\linewidth]{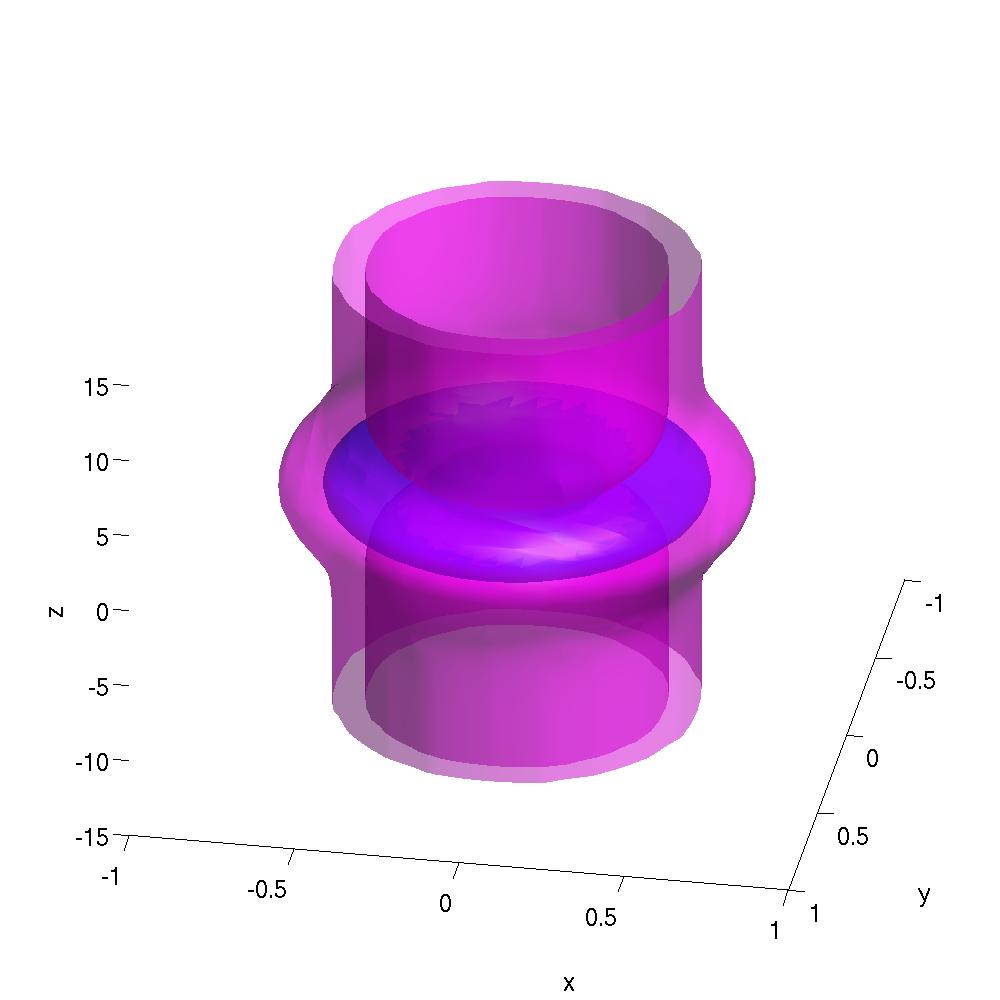}}
\subfigure[\ energy density]{
\includegraphics[width=0.4\linewidth]{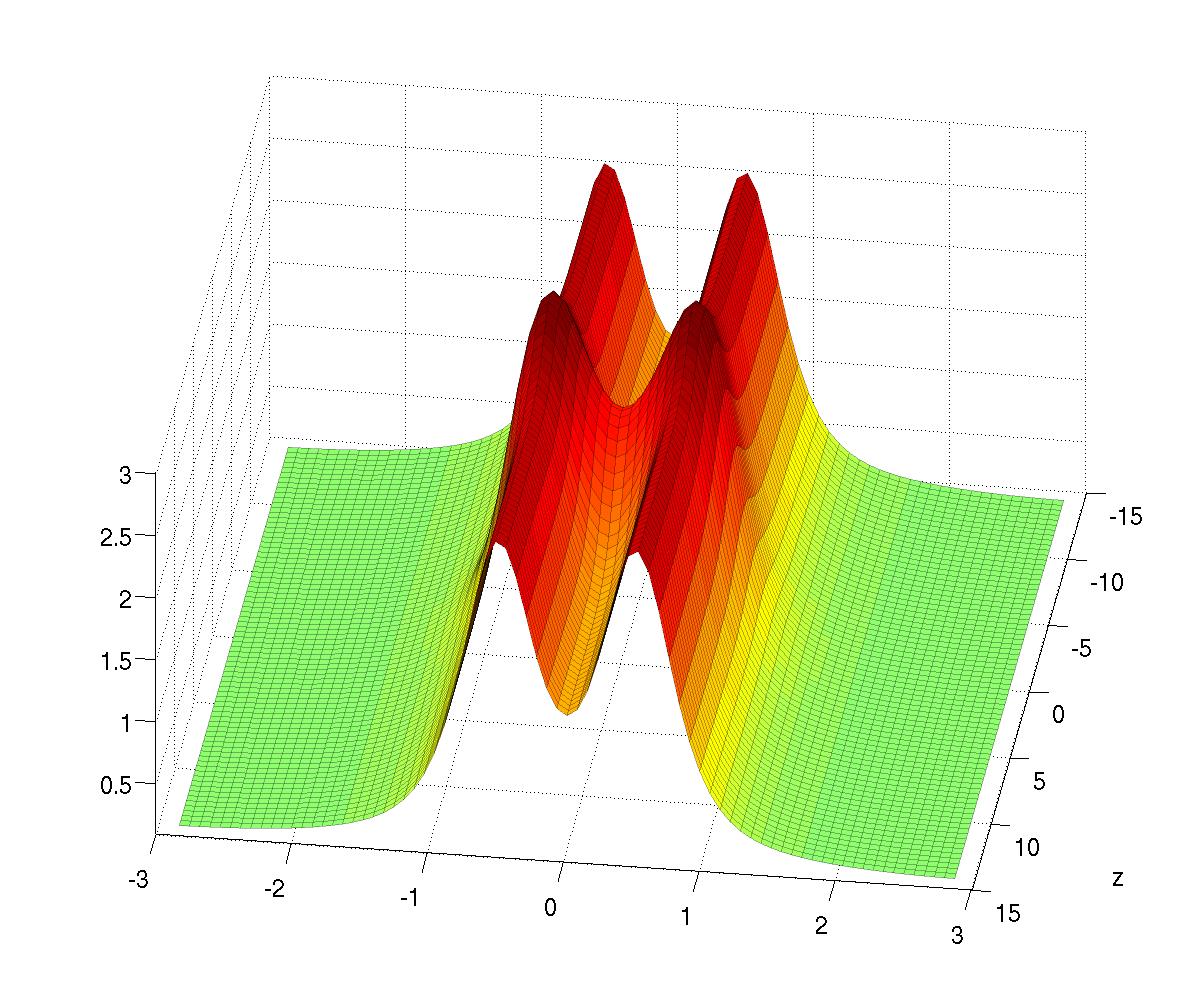}}}
\mbox{\subfigure[\ baryon charge density]{
\includegraphics[width=0.4\linewidth]{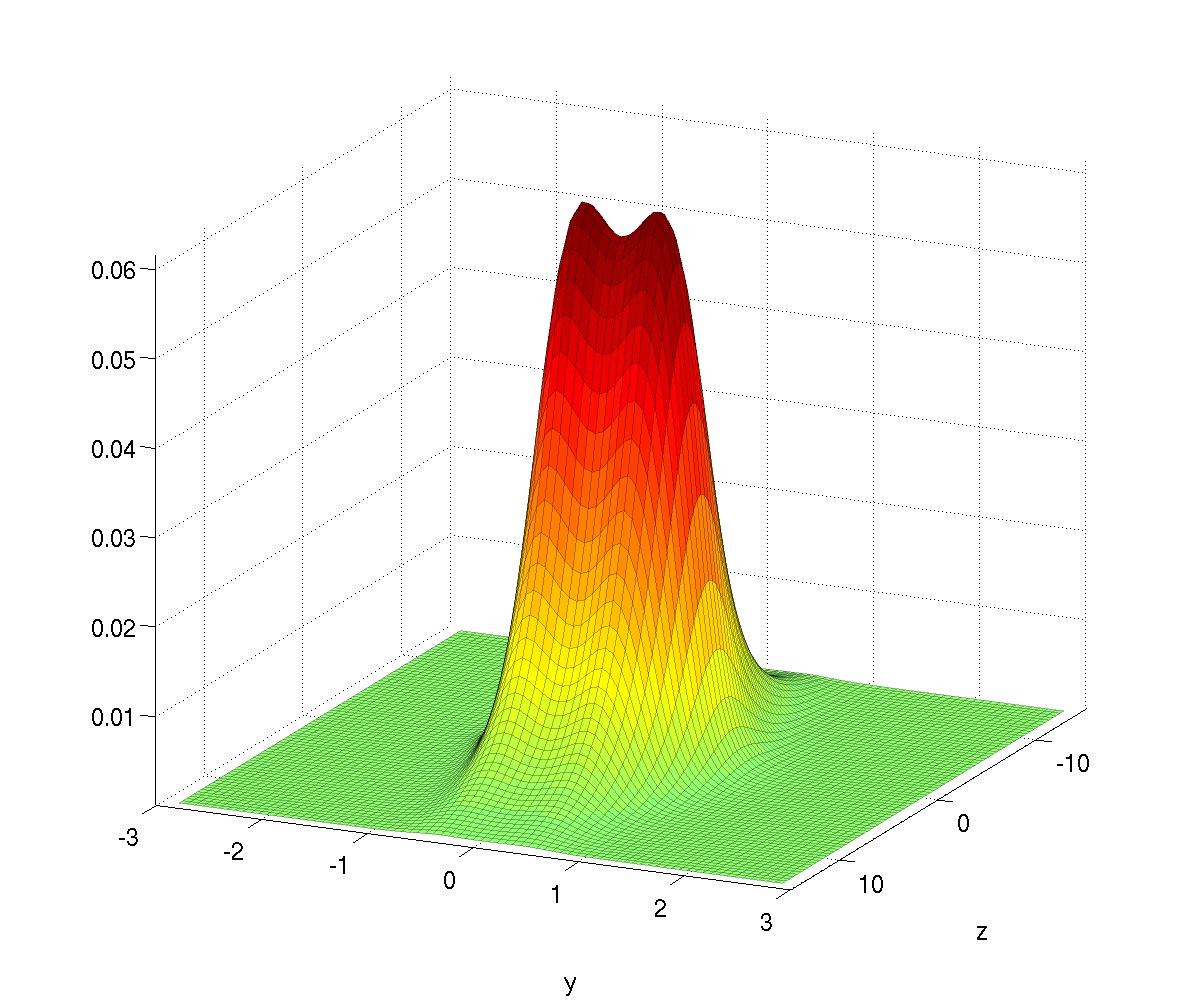}}
\subfigure[\ kink energy density]{
\includegraphics[width=0.4\linewidth]{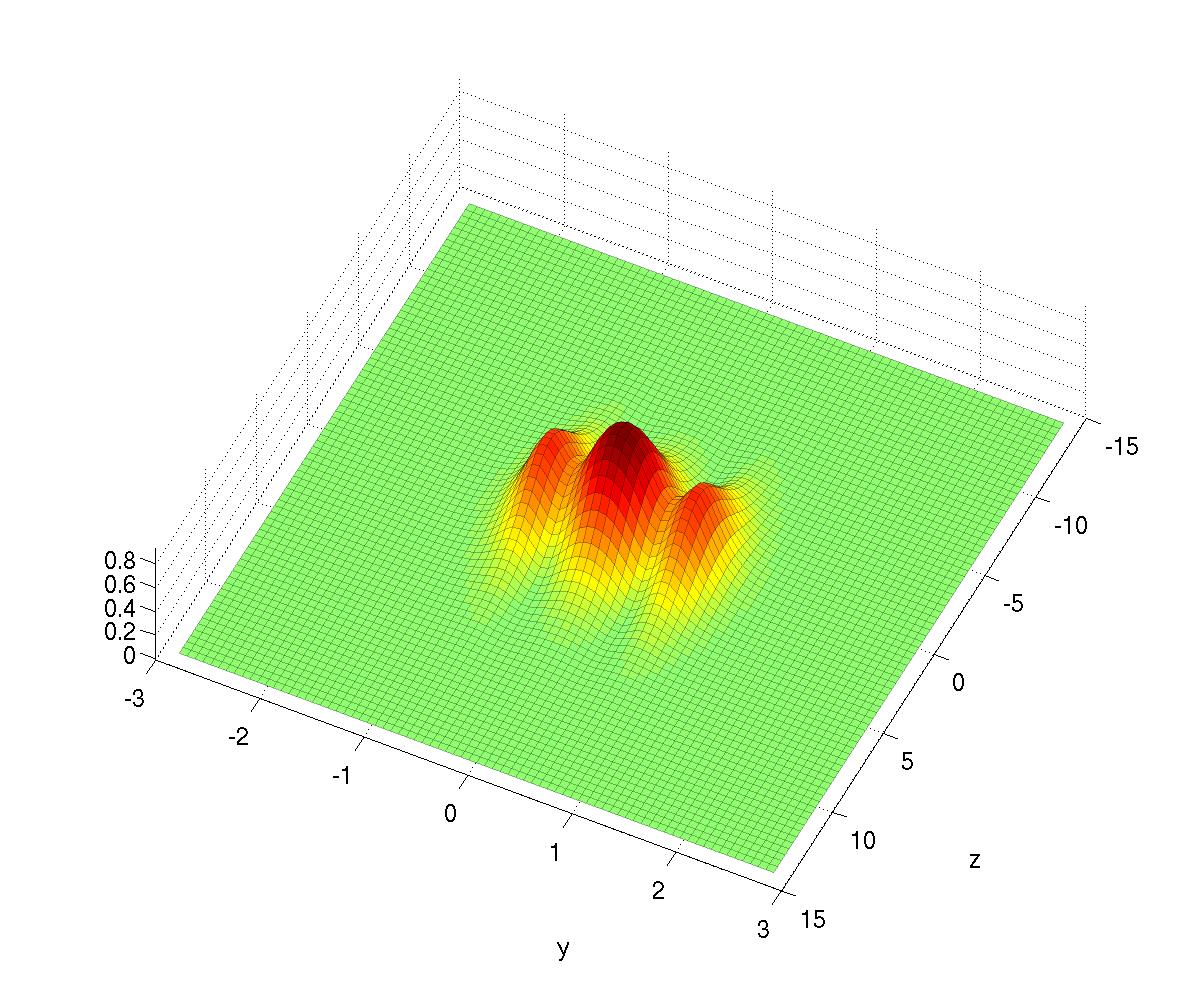}}}
\caption{The vortex with a trapped Skyrmion which is manifested as a
  sine-Gordon kink on its world-volume, in the theory with a
  sixth-order derivative term. 
  (a) 3D view of isosurfaces for the energy density and baryon charge
  density. (b) and (c) show respectively
  the energy density and baryon charge density at a $yz$-slice through
  the vortex (at $x=0$). (d) shows the energy of the kink (which is
  the total energy with the vortex energy subtracted off). Notice that
  instead of binding energy, the higher-derivative term induces
  sub-peaks on the side of the kink. 
  The calculation is done on an $81^3$ cubic lattice, with $M=3$,
  $m_3=1$ and the baryon charge is $B^{\rm numerical}=0.990$.}
\label{fig:vortexkink}
\end{center}
\end{figure}

Now we will consider a different example, namely the full sine-Gordon 
kink in the vortex theory with the sixth-order derivative term turned
on ($c_6=1$) while the Skyrme term is still off ($\kappa=0$). 
As already mentioned, it does not alter the vortex solution far away
from the kink, but it does increase the length of the kink living on
its world-volume. In Fig.~\ref{fig:vortexkink} we present this
solution and we have taken the kink mass to be $m_3=1$ and the vortex
mass $M=3$. The kink length was measured with
Eq.~\eqref{eq:kinklengthmeasure} to be $L_{\rm kink}\sim 2.44$.
The effective theory predicted this kink to be almost five times
longer than that without the sixth-order derivative term; whereas
numerically it is only about $2.7$ times longer (according to this
measure); but recall that the scaling argument is just a rough
estimate neglecting the actual integrals (or rather assuming them to
be of order one). 
We chose the kink mass in the latter vortex solution (in
Fig.~\ref{fig:vortexkink}) to be $m_3=0.22$ such that the kink should
have approximately the same length as that with a sixth-order
derivative term with $m_3=1$. Measuring this ratio numerically we get
$\sim 0.73$ which on the one hand determines the accuracy of the
estimate; but also confirms that the effective theory is qualitatively
correct.

%%%%%%%%%%%%%%%%%%%%%%
%%%%%%%%%%%%%%%%%%%%%%

\begin{figure}[!hpt]
\begin{center}
\mbox{\subfigure[\ isosurfaces]{
\includegraphics[width=0.3\linewidth]{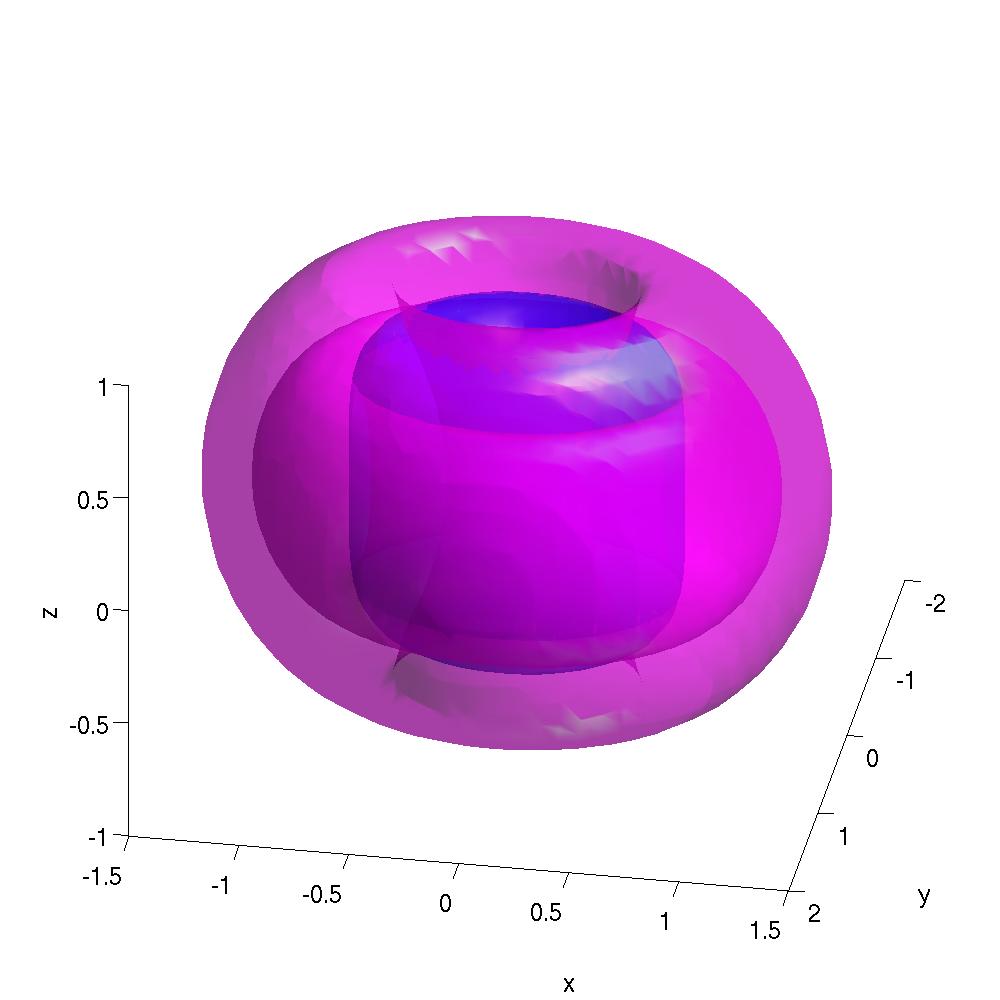}}
\subfigure[\ energy density]{
\includegraphics[width=0.3\linewidth]{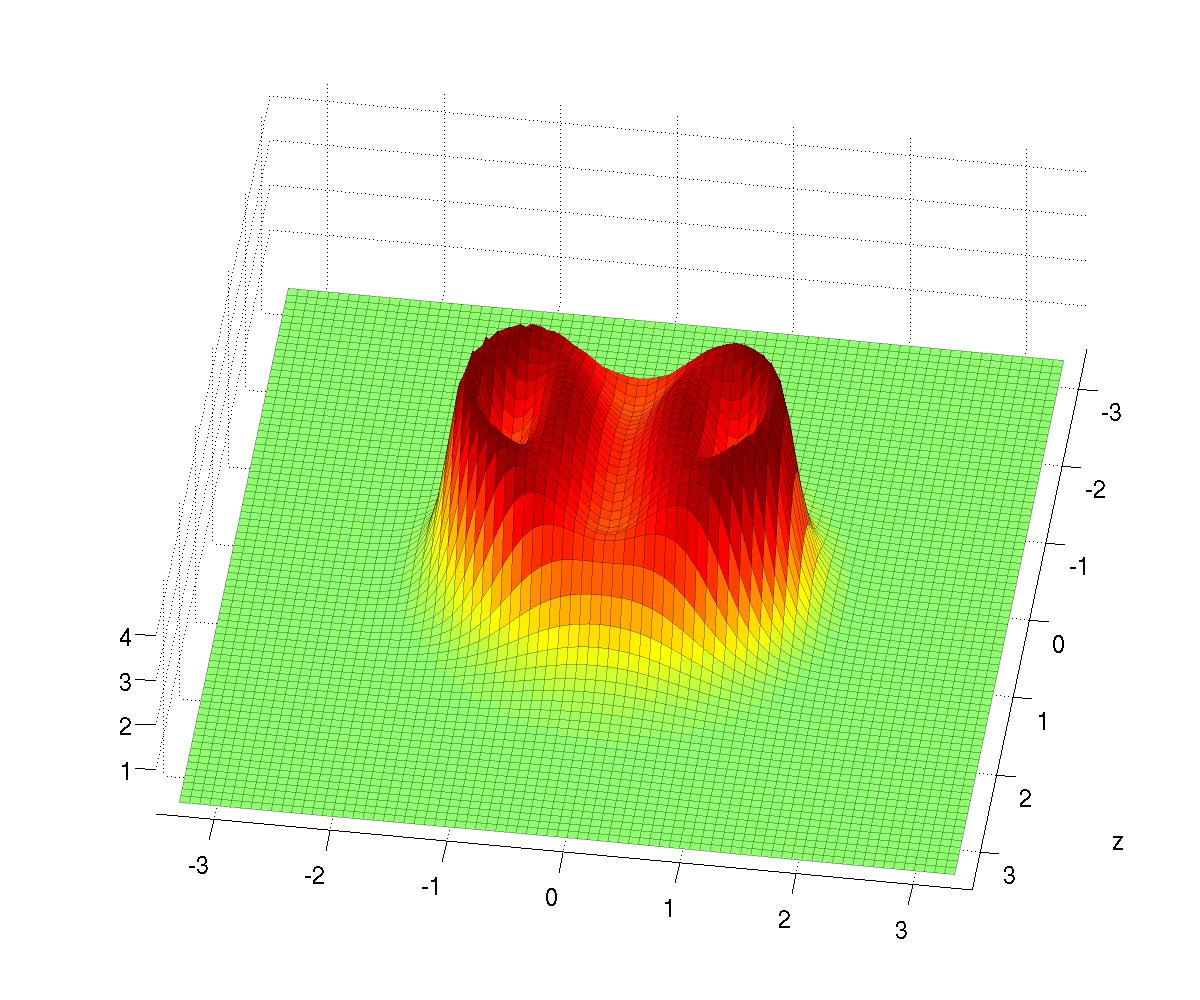}}
\subfigure[\ baryon charge density]{
\includegraphics[width=0.3\linewidth]{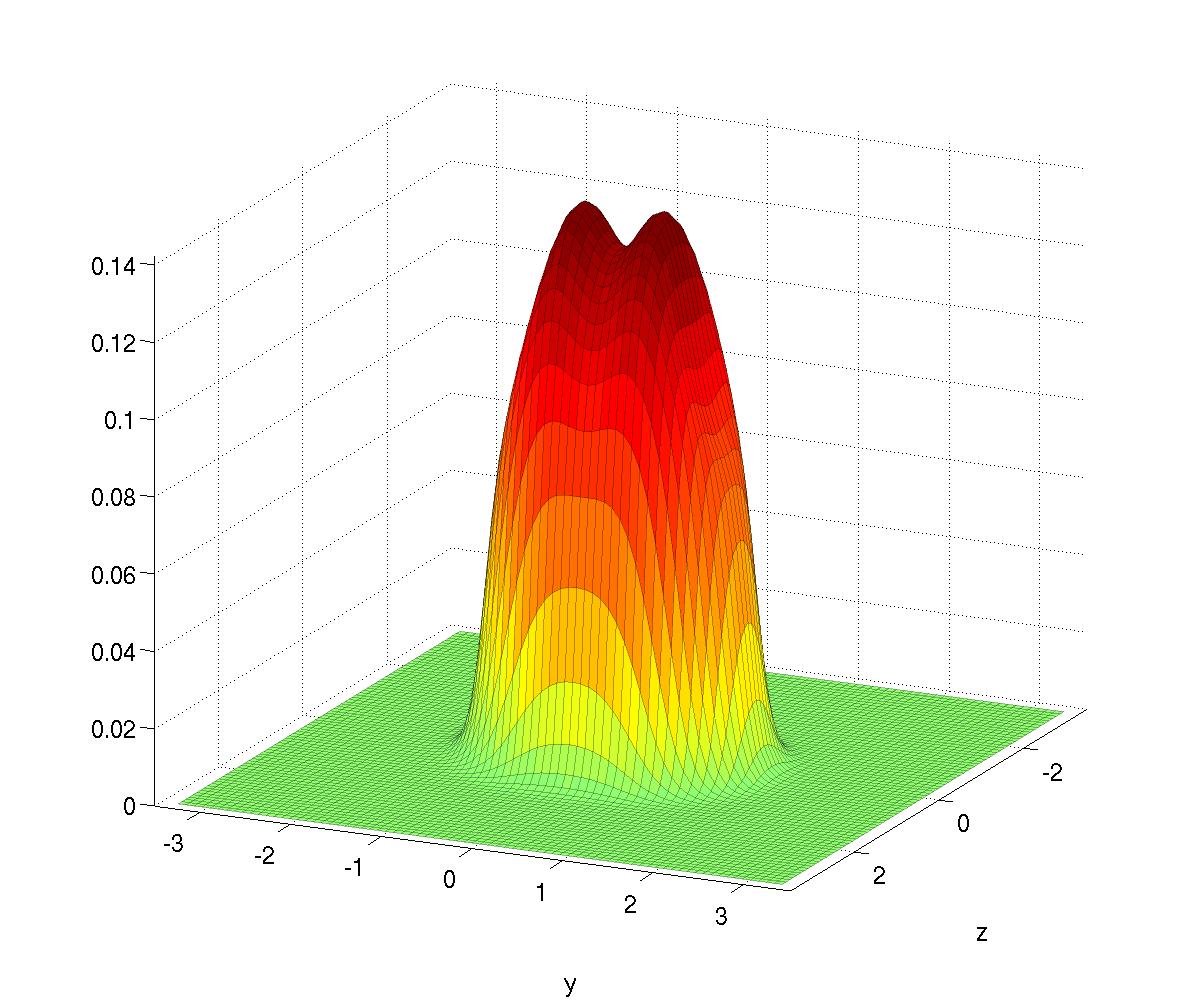}}}
\caption{The vortex compactified on a circle in the theory with a 
  sixth-order derivative term and without the potential $V_2$. 
  (a) 3D view of isosurfaces for the energy density and baryon charge
  density as in Fig.~\ref{fig:DW_SL}. (b) and (c) show respectively
  the energy density and baryon charge density at a $yz$-slice (at
  $x=0$). The calculation is done on an $81^3$ cubic lattice, with
  $M=4$ and $B^{\rm numerical}=0.99990$. }
\label{fig:torus}
\end{center}
\end{figure}

The last example we will consider, is the vortex compactified on a
circle, $S^1$, without a potential $V_2$, which has a free theory
living on its world-volume. 
Existence of this solution without angular momentum, requires
a higher-derivative term; here we will use only the sixth-order
derivative term.
In Fig.~\ref{fig:torus} is shown a numerical solution with vortex mass
$M=4$ and $c_6=1$. 
Notice that the energy density is torus-like to some extent (it has a
valley almost halfway down in energy density) but the baryon charge
density remains as a ball-like object with a dip in the energy density
at the origin.  

%%%%%%%%%%%%%%%%%%%%%%%%%%%%%%%%%%
\section{Summary and Discussion}
\label{sec:summary}

In this paper we have exhausted the possibilities (known so far) of
Skyrmions in different disguises. By trapping a Skyrmion on a domain
wall, it hosts a baby-Skyrmion while the full system has a
3-dimensional Skyrme (baryon) charge. If the domain wall is
compactified it is again a normal Skyrmion, but having its energy
distributed as in a ball-like object. Using the parameter space of the
model, it is possible to obtain a spherical shell-like object -- by
for instance having very large sixth-order derivative term, see
\cite{Gudnason:2013qba}. 
In this paper, we find the new and last piece of the puzzle, i.e.~the
Skyrmion trapped on a vortex string, which looks like a sine-Gordon
kink on the vortex world-sheet. We find the existence of this object
by an effective theory approach \cite{Gudnason:2014gla} and by
explicit numerical calculations.  The last object we find here is the
vortex compactified on a circle, which thus carries a Skyrme charge
by having a twist on its modulus. Kinks can furthermore live on
this torus-like object, but we leave such studies for future
developments.  

Let us comment on the accuracy of the comparison between the lengths
predicted from the effective theory and the actual numerical
calculation we carried out. As emphasized both here and in
\cite{Gudnason:2014gla}, the effective theory relies heavily on the
separation of scales when taking only the leading-order contribution
into account. Higher-order corrections have not yet been calculated
explicitly, although it is straightforward. However, the numerical
solutions are all done on a finite square-lattice which makes a too
large separation of scale inconvenient, i.e.~memory and run-time
consuming, which is why we have only an order 3--4 between the mass
scales in the systems studied. 

We have constructed a single sine-Gordon kink residing in a vortex in
this paper, however, it is also possible to make a sine-Gordon kink
crystal, which is described by the elliptic function ${\rm
  sn}(x)$ \footnote{See for instance Ref.~\cite{Canfora:2014aia} for
  sine-Gordon kink crystals in the context of the Skyrme model on
  $S^2\times S^1$. }.

In Tab.~\ref{table:instantons}, we have summarized the
topological incarnations of
lumps (baby-Skyrmions, or sigma-model instantons), 
Skyrmions, and Yang-Mills instantons. 
There seems to be certain relations among
the homotopy in the bulk (resultant solitons), 
the homotopy of host solitons, 
and the homotopy of world-volume solitons, 
but an exact mathematical correspondence is yet to be clarified.  
We can do the same for Hopfions (knot solitons) 
\cite{Faddeev:1996zj}; 
Hopfions can be realized as sine-Gordon kinks 
on a toroidal domain wall \cite{Kobayashi:2013bqa}. 

The BEC Skyrme model which we consider in this paper 
also admits D-brane solitons \cite{Gauntlett:2000de}, that is, 
vortices ending on a domain wall,  
since the corresponding BECs admit them
\cite{Kasamatsu:2010aq,Nitta:2012hy}.  
Therefore, this model admits various solitons with 
various codimensions: domain walls, vortices and Skyrmions, 
and their composites. 
The dynamics of these solitons remain as an interesting problem to
explore. 
For instance, Skyrmions were proposed to be created after the
annihilation of a brane and anti-brane
\cite{Nitta:2012hy,Nitta:2012kj}.

%%%%%%
\section*{Acknowledgements}

The work of M.N. is supported in part by 
a Grant-in-Aid for Scientific Research (No. 25400268) 
and by the ``Topological Quantum Phenomena'' 
Grant-in-Aid for Scientific Research 
on Innovative Areas (No. 25103720)  
from the Ministry of Education, Culture, Sports, Science and
Technology (MEXT) of Japan.  
SBG thanks Keio University (Yokohama, Japan) and Institute of Modern
Physics (Lanzhou, China) for hospitality. 

%%%%%%%%%%%%%%%%%%%%%%%%%%%%%%%%%%%%%%%%%%%%%%%%%%%%%%%%%%%%

\appendix

\section{Two-component Bose-Einstein condensates}
\label{app:BEC}
Two-component Bose-Einstein condensates 
with two wave functions $\phi_1,\phi_2$ 
have the potential term \cite{Kasamatsu:2005}
\beq
 V &=& - \mu_1 |\phi_1|^2 - \mu_2 |\phi_2|^2 
 +   \frac{g_{11}}{2} |\phi_1|^4
 +   \frac{g_{22}}{2} |\phi_2|^4 
 +   g_{12}  |\phi_1|^2 |\phi_2|^2.
\eeq
When we consider the case $g_{11}=g_{22} = g$ and $\mu_1=\mu_2=\mu$, 
the potential reads
\beq
 V &=& - \mu |\phi_1|^2 - \mu |\phi_2|^2 
 +   \frac{g}{2} |\phi_1|^4
 +   \frac{g}{2} |\phi_2|^4 
 +   g_{12}  |\phi_1|^2 |\phi_2|^2 \non
&=& \frac{g}{2} \left(|\phi_1|^2 + |\phi_2|^2 - v^2\right)^2 
    + m^2  |\phi_1|^2 |\phi_2|^2 + {\rm const.}
\eeq
with 
\beq
v^2 \equiv \frac{\mu}{g},\quad 
\frac{1}{2}m^2 \equiv g_{12} - g.
\eeq
We consider the strong coupling limit 
\beq
 g , g_{12} \to \infty,  \qquad m^2 = {\rm fixed},
\eeq
which yields the $O(4)$ nonlinear sigma model 
with the constraint $|\phi_1|^2 + |\phi_2|^2 = v^2$, 
whose target space is $S^3 \simeq SU(2)$. 
The potential is
\beq
 V = \frac{1}{2}m^2 |\phi_1|^2 |\phi_2|^2 + {\rm const.}
\eeq
which is of the form considered in this paper. 

%%%%%%%%%%%%%%%%%%%%%%%%
\section{Half a Skyrmion trapped on a vortex}\label{app:half}

\begin{figure}[!tbh]
\begin{center}
\mbox{\subfigure[\ isosurfaces]{
\includegraphics[width=0.4\linewidth]{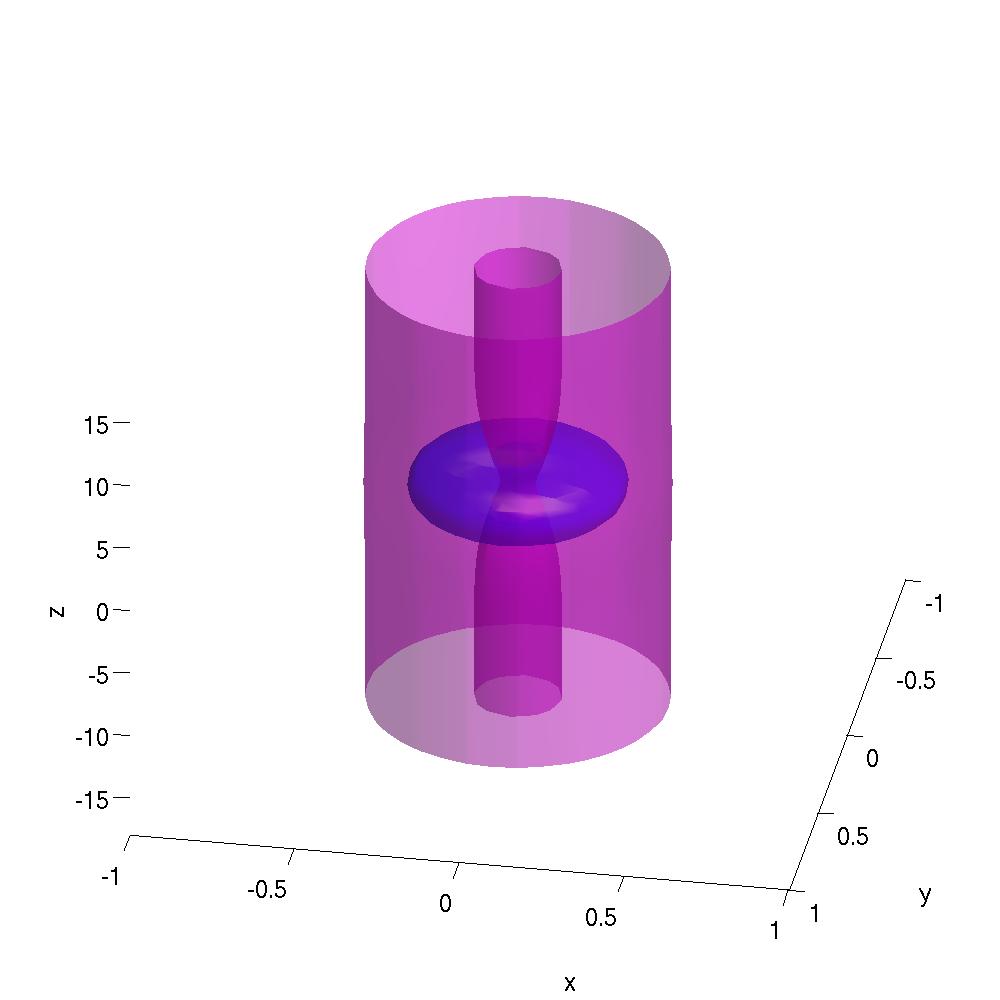}}
\subfigure[\ energy density]{
\includegraphics[width=0.4\linewidth]{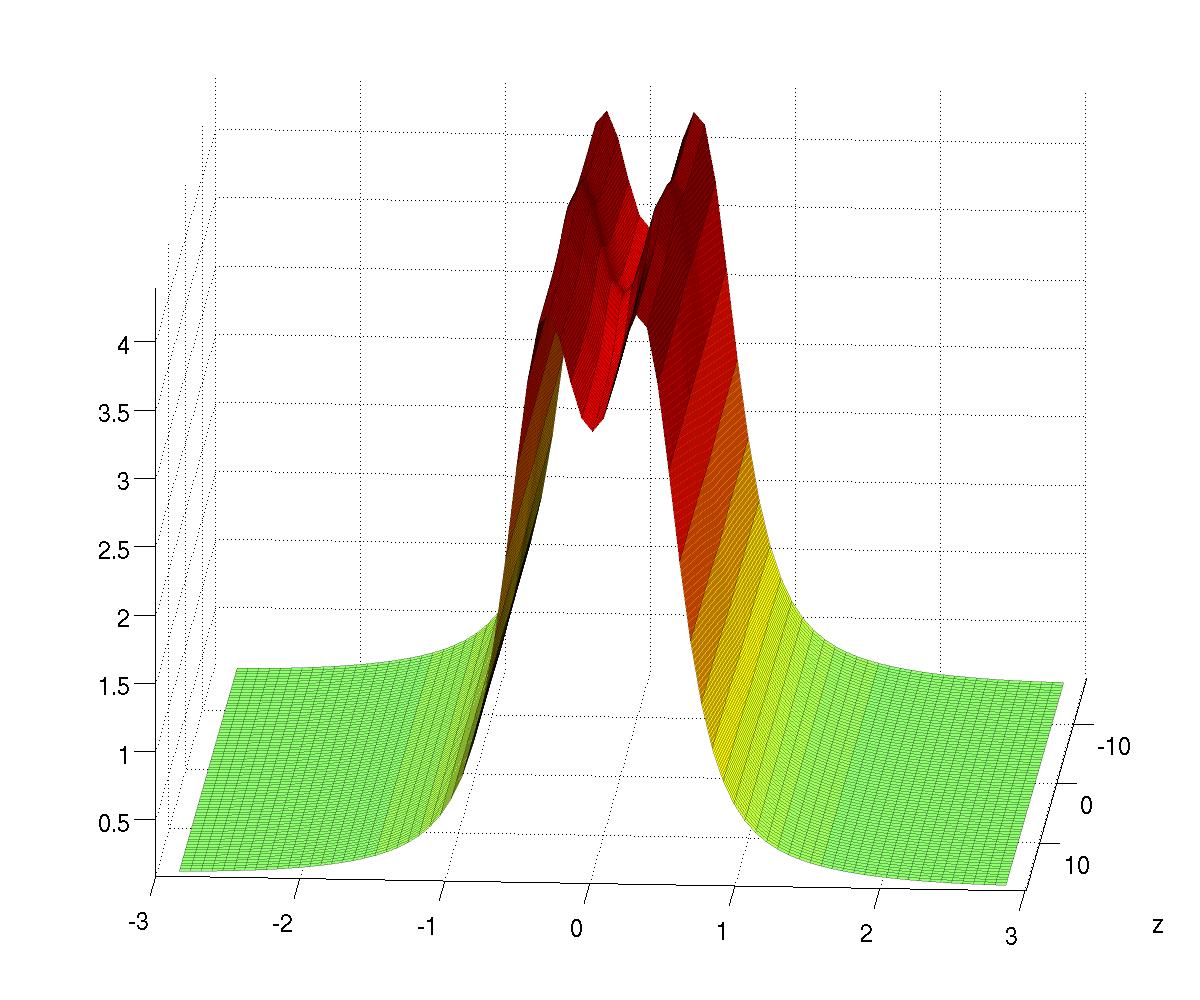}}}
\mbox{\subfigure[\ baryon charge density]{
\includegraphics[width=0.4\linewidth]{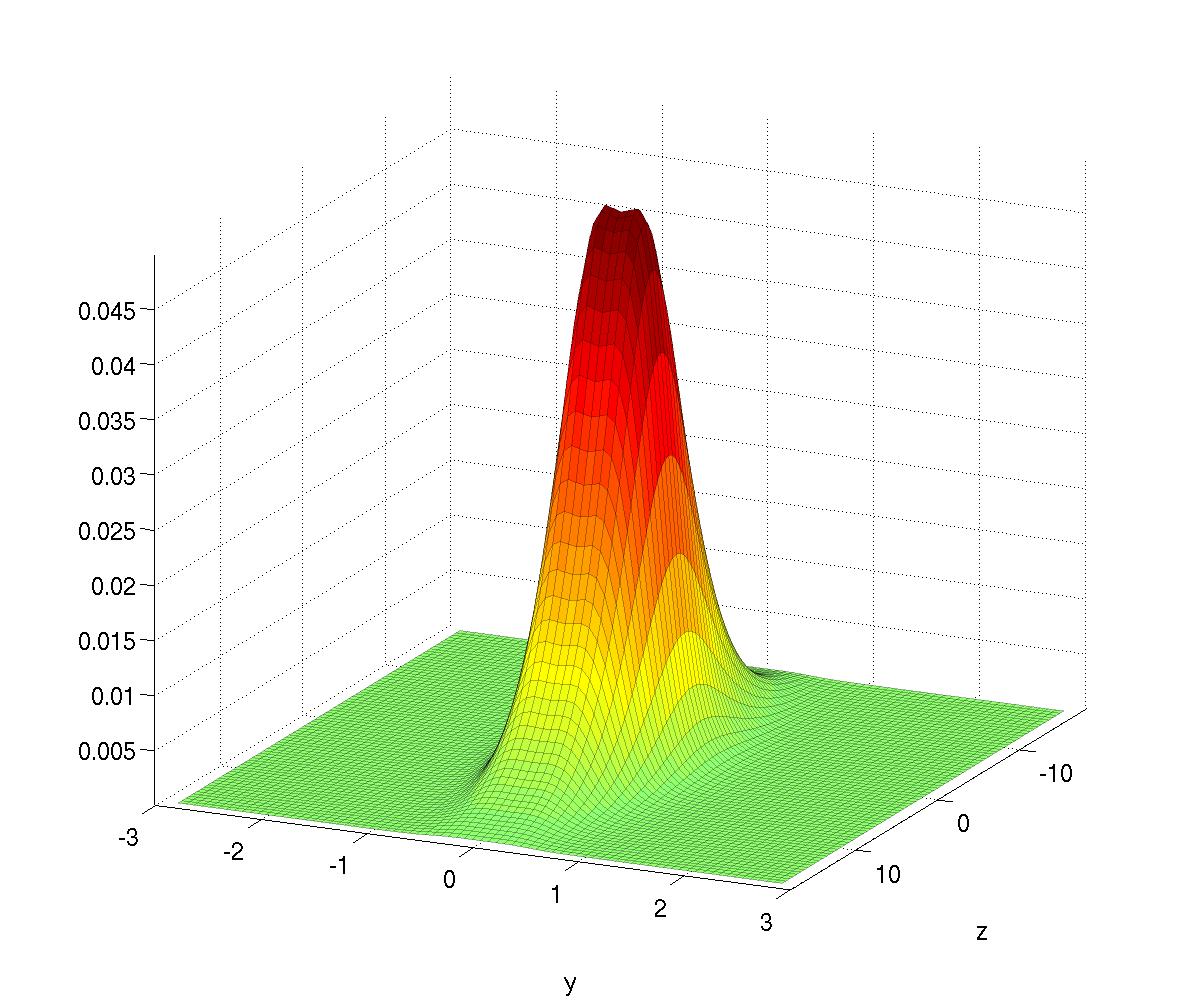}}
\subfigure[\ kink energy density]{
\includegraphics[width=0.4\linewidth]{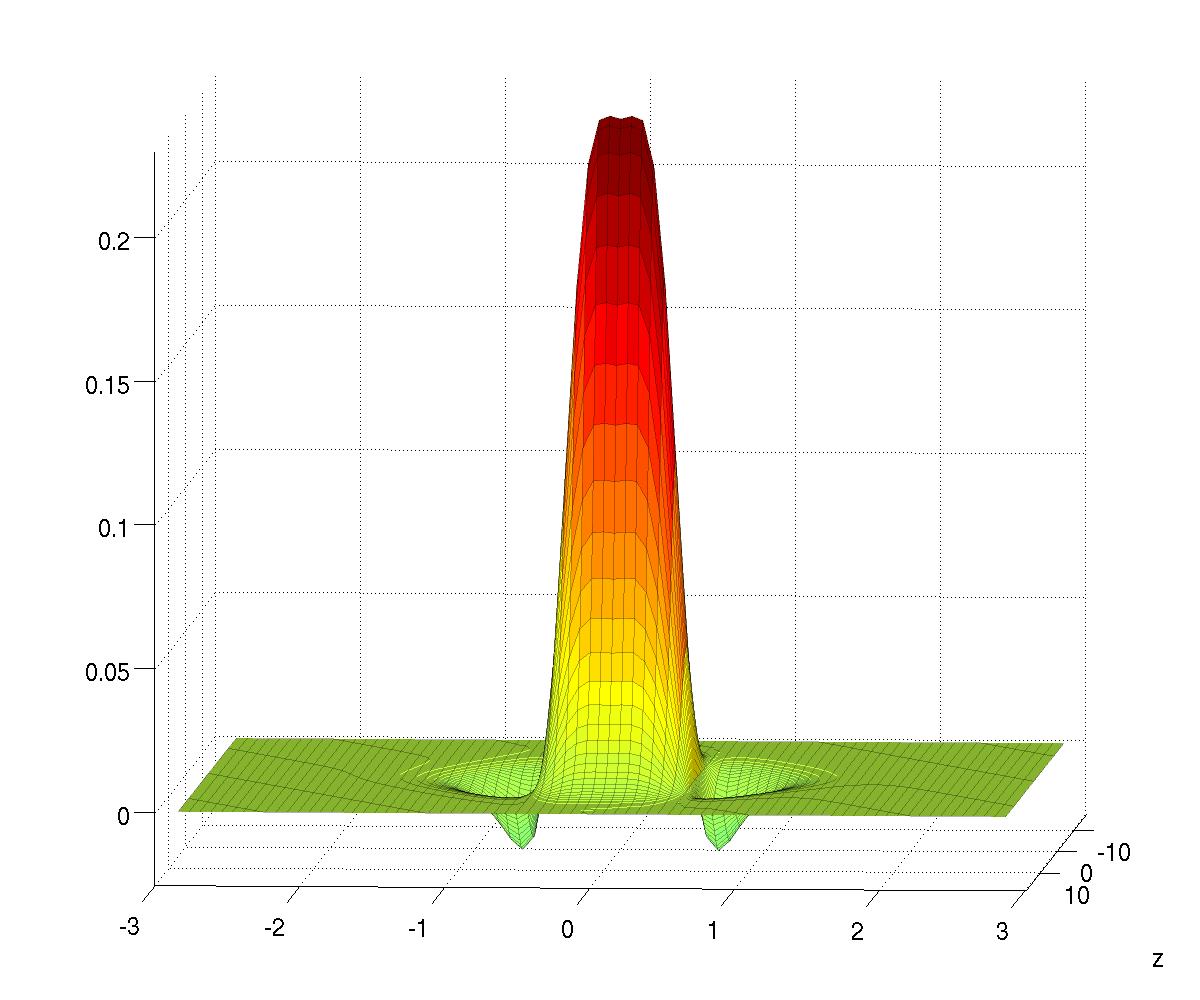}}}
\caption{The vortex with a trapped half-Skyrmion which is manifested
  as ``half'' a sine-Gordon kink on its world-volume, in the theory
  with no higher-derivative terms. 
  (a) 3D view of isosurfaces for the energy density and baryon charge
  density. (b) and (c) show respectively
  the energy density and baryon charge density at a $yz$-slice through
  the vortex (at $x=0$). (d) shows the energy of the kink (which is
  the total energy with the vortex energy subtracted off). Notice the
  negative dips on each side of the peak in the kink energy; we
  interpret those as binding energy. 
  The calculation is done on an $81^3$ cubic lattice, with $M=3$,
  $m_3=0.22$ and the baryon charge is $B^{\rm numerical}=0.497$. }
\label{fig:vortexhalfkink_ko}
\end{center}
\end{figure}

As promised earlier in the paper, half a Skyrmion can be manifested by
making a sine-Gordon kink with the potential $V_2$ and setting
$a_3=2$, such that only ``half'' a twist of the $U(1)$ modulus is
needed, resulting in a Skyrmion with half a unit of baryon charge. 
The solution is shown for the vortex theory without higher-derivative
terms, vortex mass $M=3$, kink mass $m_3=0.22$ and is shown in
Fig.~\ref{fig:vortexhalfkink_ko}. As can be seen from the figures, the 
half-Skyrmion is qualitatively the same as the full Skyrmion, except
for possessing only half the baryon charge. 

%%%%%%%%%%%%%%%%%%%%%%%%%%%%%%%%%%%%%%%%%%%%%%

\end{document}